# A Hybrid Agent-Based and System Dynamics Framework for Modelling Project Execution and Technology Maturity in Early-Stage R&D

Robson Wilson Silva Pessoa[a], Marie Hahn Naess[a], Julia Carolina Bijos[a], Carine Menezes Rebello[a], Danilo Colombo[b], Leizer Schnitman[c] and Idelfonso B. R. Nogueira[a,*]

[a]*Department of Chemical Engineering, Norwegian University of Science and Technology, Trondheim, 793101, Norway*
[b]*CENPES, Petrobras R&D Center, Av. Horácio Macedo 950,Cid. Universitária, Ilha do Fundão, Salvador, Brazil*
[c]*Postgraduate Program in Mechatronics, Federal University of Bahia, Polytechnic School, R. Prof. Aristídes Novis, 2, Federação, Salvador, 40210-630, Brazil*

ABSTRACT

This paper presents a hybrid approach to predict the evolution of technological maturity of R&D projects, using the context of the oil and gas (O&G) sector as an example. Integrating System Dynamics (SD) and Agent-based Modelling (ABM) enables the proposed multi-level framework to capture uncertainties inherent to R&D projects, including work effort, team size, and project duration, all of which influence technological progress. Although AB–SD hybrid models are well established in other fields, their application in R&D contexts remains limited. The AB–SD model combines system-level feedback structures governing work phases, rework cycles, and project duration with the explicit representation of decentralised agents (e.g., team members, tasks, and controllers) whose interactions drive emergent project dynamics. A base-case scenario was developed to analyse the structural dynamics of early-stage innovation projects, simulating 15 parallel tasks over 156 weeks. In a comparative scenario with sequential task execution, the model showed an 88% reduction in rework duration relative to the base case. The second scenario evaluated mixed parallel–sequential task structures under varying team sizes. In the parallel configuration, simulation results indicated that increasing team size reduced overall project duration and improved task completion rates, with optimal performance achieved for teams of four to five members. These outcomes are consistent with empirical observations in R&D project management, where moderate team expansion enhances coordination efficiency without incurring communication overhead. However, as widely recognised in empirical studies, a substantial increase in team size does not necessarily translate into higher completion rates, as excessive team growth often introduces communication complexity and management delays. Overall, the model outputs and the proposed modelling framework are well aligned with expert understanding in the field, confirming their validity as a quantitative tool for analysing resource allocation, task scheduling efficiency, and technology maturity progression.

## 1. Introduction

Over the past decade, energy-transition policies have increasingly driven industry to enhance both the economic and environmental performance of its processes (Yang et al., 2024). This transformation is shaped by technological advancements, as well as the evolving organisation of work and the coordination of innovation activities. Within this systemic view, Hekkert and Negro (2009) identify seven key functions of technological innovation systems—entrepreneurial activities, knowledge development, knowledge diffusion through networks, direction of research, market formation, mobilisation of human and financial resources, and the creation of legitimacy to overcome resistance to change—which together underpin the management of technological and institutional transitions.

Building on this framework, Klessova et al. (2020) examined how knowledge development relates to the mobilisation and allocation of resources in R&D environments. Through data analysis and interviews across European Union research and innovation programmes encompassing projects at different levels of technology maturity, they identified organisational structures that link project management practices with mechanisms of knowledge integration.

---

*Corresponding author

✉ robson.w.s.pessoa@ntnu.no (R.W.S. Pessoa); idelfonso.b.d.r.nogueira@ntnu.no (I.B.R. Nogueira)
ORCID(s): 0000-0003-1603-1453 (R.W.S. Pessoa)

Their findings highlight the intrinsically multi-actor nature of R&D coordination, where diverse participants interact to align knowledge flows, resources, and objectives across the innovation lifecycle.

Large energy companies employ diverse implementation strategies and foster interactions across the functions of technological innovation systems, establishing organisational arrangements that strengthen engagement with scientific and technological institutions. Although Hekkert and Negro (2009) note that the relationship between project maturity and these system functions remains insufficiently understood, the oil and gas sector offers a particularly relevant example. This industry faces the dual imperative of maintaining a reliable energy supply while accelerating the adoption of low-carbon technologies and compliance with environmental regulations. The resulting pressure to innovate—rapidly improving process efficiency, emissions control, and energy integration—has led firms to increasingly rely on structured R&D frameworks and cross-sectoral collaborations that link science, technology, and industrial practice. In this context, both qualitative and quantitative data can be leveraged to model projects governed by the emergent dynamics of these innovation functions, enabling more systematic management of technology development under transition constraints.

Given that R&D environments integrate complex technical, organisational, and market dimensions, a broader analytical approach is required to represent technological progress comprehensively. Structural modelling methods have proven particularly effective for this purpose, as they enable the explicit representation of cause–effect relationships among interdependent variables, thereby providing a systemic understanding of technological innovation processes (Daim et al., 2006). In their study, Daim et al. (2006) demonstrated how System Dynamics can be combined with bibliometric and patent analyses to forecast emerging technologies, bridging technical, organisational, and institutional perspectives within the innovation ecosystem.

Within this methodological landscape, System Dynamics plays a pivotal role by allowing the simultaneous modelling of technological, financial, policy, project management, and sustainability dimensions. Its capacity to represent feedback structures and non-linear behaviours through causal loops and stock-and-flow diagrams supports both qualitative and quantitative analysis of innovation systems (Shafieezadeh et al., 2020; Borshchev and Filippov, 2004; Chang et al., 2022). Borshchev and Filippov (2004) highlighted the integration of System Dynamics with Agent-Based approaches to capture interactions between entities, while Shafieezadeh et al. (2020) and Chang et al. (2022) demonstrated its application to project performance assessment and policy evaluation in engineering contexts. Together, these studies reinforce the suitability of System Dynamics for analysing the complex, multi-actor processes that underpin technological innovation and the ongoing transformation of the energy sector.

Despite these advances, there is still no established application of such hybrid modelling approaches in modelling early-stage industrial R&D projects. Modelling in this setting involves specific challenges, as it must account for the high uncertainty, multi-level decision making, and asynchronous task execution that characterise innovation processes, while capturing their interaction with technology maturity progression. Integrating the micro-level behaviour of agents, such as teams and tasks, with the macro-level feedback structures that govern coordination, rework, and resource allocation remains a central challenge. The model must represent these interdependencies with sufficient granularity to reflect operational reality, while retaining transparency and interpretability for decision support.

The integrated analysis presented by Daim et al. (2006) demonstrated the potential of System Dynamics to improve the understanding of the elements shaping alternative scenarios in the diffusion of emerging technologies. Their work identified government policies, cost structures, and environmental aspects as key drivers influencing the implementation and market dissemination of new technologies. Similarly, Eghbali et al. (2022) confirmed that government intervention plays a central role in strengthening partnerships between firms and green startups, whether through financial subsidies, the creation of incubators, or the establishment of regulatory frameworks designed to reduce the risks associated with adopting green technological innovations. In both studies, technology maturity emerged as a critical determinant of adoption risk. Daim et al. (2006) approached this by using bibliometric and patent data to estimate maturity levels, while Eghbali et al. (2022) formalised maturity as a parameter in an evolutionary game model. Although such representations offer valuable insights, they tend to remain case-specific and may limit the generalisation of results across different technological and organisational settings.

The concept of Technology Readiness Level (TRL) provides a systematic alternative for representing technological maturity. Originally developed by NASA as a means to evaluate the readiness of space technologies, the TRL scale quantifies the progression from basic principles to fully operational systems, reflecting the balance between technological capability and the financial and organisational resources supporting development. Over time, the framework has been adapted to a wide range of industries, including the energy sector, where it has been applied to assess technology evolution, investment risk, and deployment readiness (Commission et al., 2017; Buchner et al., 2019).

Each TRL stage captures distinct aspects of technological advancement, from conceptual research to demonstration and commercialisation.

Despite its widespread adoption, a gap persists in the literature regarding the explicit integration of the seven functions of technological innovation systems with project maturity metrics such as TRL, particularly in the context of industrial energy R&D projects. Incorporating TRL into dynamic modelling remains a methodological challenge, as it requires representing the complex, multi-level structure of technological progression and the interdependencies that exist across readiness stages. Addressing this gap is essential to provide managers and policymakers with operational tools for optimising project portfolios, aligning innovation strategies with systemic functions, and enhancing the effectiveness of technology development under energy transition constraints.

Building on the recognised importance of TRL as a maturity assessment framework, several studies have sought to understand its practical implications and methodological extensions. Lavoie and Daim (2017) conducted a qualitative study using Grounded Theory and interviews with researchers and practitioners to examine the role of TRLs in R&D management. Their findings confirmed that TRLs are widely regarded as essential instruments for assessing technological maturity, largely due to their value as a standardised and communicable reference across projects. However, the study also identified several limitations, including subjectivity in assessments, the lack of integration with broader management systems, and persistent ambiguities in TRL definitions. Although some participants referred to complementary frameworks such as PMBOK, OPM3, and CMMI, these were considered inadequate substitutes for TRLs, reinforcing the need for approaches that better integrate maturity assessment within systemic models of innovation.

Addressing this methodological challenge requires developing a coherent framework capable of connecting simple causal relationships into an integrated chain of decisions that reflects the systemic behaviour of institutional functions. In this regard, Calderon-Tellez et al. (2024) explored decision-making support mechanisms based on the dimensions of innovation and sustainability, emphasising the use of causal loop structures to represent feedback and interdependencies. Nevertheless, the innovation dimensions considered in their framework did not explicitly incorporate classical quantitative indicators such as TRL, which are central to R&D evaluation. Complementarily, Kenley et al. (2024) proposed a statistical method to estimate the probability of TRL evolution using historical data from NASA and the U.S. Department of Energy, while Pujotomo et al. (2025) modelled the innovation ecosystem as an interacting network of projects positioned at different TRL levels. The latter demonstrated that analysing these interconnections can reveal pathways to accelerate technology commercialisation and to align research progress with market deployment objectives.

Building on these insights, system analysis has advanced the understanding of factors influencing technological progress and innovation performance. Nevertheless, the inherent complexity of such systems is not fully captured by conventional SD models, which primarily operate at the aggregate level. This limitation has motivated the adoption of complementary approaches that enable a multi-level representation of innovation ecosystems. Micro-level modelling, in particular, allows the explicit representation of agents and their interactions, capturing heterogeneity, nonlinearity, and spatial relationships across time (Langarudi et al., 2021). Although hybrid frameworks that combine SD with Agent-Based Modelling (ABM) have demonstrated significant potential in other domains, their application to the planning, monitoring, and management of industrial R&D processes remains scarce in the literature.

The present study addresses this gap by introducing a hybrid, dynamic, and predictive modelling framework that integrates the structural perspective of SD with the behavioural richness of multi-agent systems. The proposed model captures the evolution of R&D projects by linking functional aspects such as rework cycles, coordination loops, and schedule dependencies with the characteristics of decentralised agents, including team members, tasks, and controllers. A key feature of the framework is its capacity to incorporate TRL prediction, allowing the estimation of technological maturity progression as an endogenous outcome of project dynamics. Through this integration, the model represents emergent behaviours that influence project performance and technology advancement over time. BThe framework provides a practical decision-support tool that allows energy and industrial firms to relate project dynamics to technology maturity evolution, supporting more informed decisions on resource allocation, scheduling, and strategic portfolio management under energy transition constraints. The proposed methodology is tested in the context of the oil and gas sector. In this environment, firms are required to optimise R&D project portfolios, enhance collaboration between research units and production systems, and reduce time-to-deployment of new technologies. Applying the model to this setting demonstrates its ability to reproduce the dynamics of early-stage innovation projects, assess the impact of team configuration and scheduling strategies, and predict the evolution of TRL as a function of project management practices. This case application illustrates the framework's potential as both an analytical and operational tool for guiding technology development in complex industrial ecosystems undergoing energy transition.

## 2. Methodology

The methodology developed in this study establishes a hybrid System Dynamics and Agent-Based Modelling framework designed to represent the complex dynamics of R&D project execution and its relationship to technology maturity progression. The approach follows a multi-layered structure that integrates macro-level feedback mechanisms with micro-level behavioural interactions. The first stage involves constructing the SD model, which captures the aggregate evolution of work effort through stocks, flows, and feedback loops representing quality management, rework, and approval processes. The second stage introduces the ABM component, which simulates decentralised decision-making, task execution, and learning at the agent level. These two modelling paradigms are then coupled mathematically through a bidirectional data exchange architecture that enables continuous interaction between individual task states and system-level indicators.

The subsequent sections detail each component and its role within the hybrid architecture. Section 2.1 presents the development of the SD model and its stock–flow formulation for R&D contexts. Section 2.2 introduces the ABM layer and explains its integration with the SD framework. Section 2.3 formalises the coupling logic, showing how information is exchanged between layers through aggregation and feedback functions. Section 2.4 extends the model to estimate technological maturity using a probabilistic TRL formulation derived from simulation results. Section 2.5 describes the Simulation Process Manager (SPM), the integration engine that synchronises discrete agent behaviour with continuous SD dynamics. Finally, the concluding subsections address parameterisation, validation, and the guiding hypotheses that underpin this first implementation of the hybrid framework. Together, these methodological steps provide a structured foundation for analysing how micro-level task behaviour, macro-level system feedback, and maturity progression co-evolve within R&D projects.

### 2.1. System Dynamics Model
#### 2.1.1. System Dynamics: quantitative and qualitative aspects

System Dynamics is a methodological approach used to analyse the evolution of complex systems through endogenous feedback loops, stocks (accumulations), flows, and time delays (Sterman, 2002; Forrester, 2007). Its main strength lies in combining conceptual understanding with empirical evidence, effectively bridging qualitative reasoning and quantitative analysis. The approach allows for the identification and quantification of cause-and-effect relationships among variables, enabling the evaluation of complex scenarios and the representation of dynamic, nonlinear behaviours across interdependent factors. This capability makes SD particularly suitable for modelling R&D projects, where innovation processes evolve under uncertainty related to funding availability, team performance, policy influences, or incomplete data that can constrain technology development. In such contexts, SD supports strategic decision making, allowing project managers and policymakers to test scenarios, assess long-term implications, and understand how technology development trajectories may unfold over time (Shafieezadeh et al., 2020).

In SD models, stocks represent measurable quantities within the system at a specific moment, such as accumulated knowledge in an R&D project, completed activities, or available financial and human resources (Borshchev and Filippov, 2004). Flows define the rates of change in these stocks, including investment inflow, work output, or reductions caused by delays in regulatory reviews (Sterman, 2002). The relationship between stocks and flows is expressed mathematically through integration over time. The value of a stock at time $t$ results from the cumulative effect of its inflows and outflows, starting from an initial value $Stock(t_0)$, as defined in the following equation:

$$\text{Stock}(t) = \int_{t_0}^{t} [\,\text{Inflow}(s) - \text{Outflow}(s)\,]\,ds + \text{Stock}(t_0) \tag{1}$$

where Inflow($s$) represents the rate at which a quantity enters the stock and Outflow($s$) denotes the rate at which the quantity exits the stock. In discrete simulation, this continuous formulation is approximated iteratively using time steps $dt$

$$\text{Stock}(t + \Delta t) = \text{Stock}(t) + \sum_{i} \text{Inflow}_i(t)\Delta t - \sum_{j} \text{Outflow}_j(t)\Delta t \tag{2}$$

Stock–flow relationships can also represent rework cycles, a recurring process often perceived as inefficiency. In R&D projects, however, rework is a critical element of iterative learning. Uncertainty is reduced as new insights emerge, making rework both a corrective and a discovery mechanism. For instance, during the development of technology, an initial prototype may generate knowledge that necessitates design refinement. Such cycles continue until a satisfactory level of maturity is achieved, fostering innovation through feedback and continuous adjustment.

Shafieezadeh et al. (2020) illustrated this process with four key stock variables—Work to do, Work awaiting quality assessment, Work under client review, and Work approved. The associated flows represented work introduction, approval, and rework cycles, showing how progress toward completion gradually reduces total workload. Workforce behaviour was modelled as a function of team size and productivity, incorporating motivation, fatigue, and schedule pressure as factors affecting quality.

Chang et al. (2022) expanded this view, emphasising the project manager's role in shaping performance outcomes in engineering projects within government agencies. Their model captured workforce dynamics through hiring and termination flows and represented quality management through human error and rework as capacity and capability correction factors. Additional variables, including workload, knowledge growth, and project maturity, demonstrated the pivotal influence of leadership on project performance and learning.

Pujotomo et al. (2025) further extended the application of SD toward the commercialisation of research-based technologies. Their model represented TRL evolution through stocks and flows controlled by variables such as funding availability, development speed, team capacity, and rework. The results highlighted complex interdependencies between funding, team allocation, rework intensity, and TRL progression, showing that R&D trajectories depend on both resource availability and market readiness.

Causal Loop Diagrams (CLDs) complement this quantitative representation by capturing the feedback structures that drive behaviour in complex systems. CLDs depict causal relations indicating how a change in one variable influences others (Sterman, 2002). They are widely used to conceptualise dynamic behaviour, support expert model building, and facilitate systemic reasoning (Calderon-Tellez et al., 2024). Within product development and innovation, Calderon-Tellez et al. (2024) identified four core variables—social innovation, project innovation, investment in R&D, and research activities—and demonstrated that innovation adoption reduces costs while improving product benefits. Shafieezadeh et al. (2020) also examined similar interactions in oil and gas projects, revealing that human error and schedule pressure strongly influence quality and reliability, while financial feedback loops connect resource allocation, productivity, and project value.

At a broader level, Oliveira and Negro (2019) analysed biogas Technology Innovation Systems (TIS), showing that early TIS development is driven by external factors such as regulatory maturity and resource availability, whereas consolidation depends on internal mechanisms, including knowledge-sharing networks and coordinated policy. Similarly, Pujotomo et al. (2025) demonstrated, through CLDs, that funding and user acquisition form reinforcing loops accelerating technology diffusion, while prolonged regulatory phases and frequent rework act as balancing loops, slowing TRL advancement.

Managing R&D projects in the oil and gas industry adds further complexity due to high investment costs, numerous interdependent tasks, and the need for specialised expertise. These conditions underscore the importance of robust methods that can effectively explore system responses to change and prevent inefficiencies. The works of Pujotomo et al. (2025) and Shafieezadeh et al. (2020) confirm that SD provides a suitable framework for scenario analysis and systemic decision-making under uncertainty.

Building upon this foundation, the present study employs SD as the core structure for representing feedback mechanisms in R&D project evolution. The proposed framework extends traditional SD models by integrating a multi-agent layer and a TRL prediction component, enabling the simulation of both collective and individual behaviours that influence technology development. This integration advances the modelling of innovation processes in complex industrial ecosystems, offering a more comprehensive understanding of how feedback, learning, and resource allocation shape technological maturity over time.

### 2.1.2. A Novel Approach Based on System Dynamics for R&D

This study proposes a novel model that captures the system-level dynamics of R&D project execution through an SD submodel that simulates the continuous and cumulative evolution of work across key project stages. The formulation represents a simplified but representative structure of R&D workflows, reflecting the influence of rework, quality management, and approval processes on project progress and technological maturity. As detailed in Section 2.1.1,

System Dynamics provides a robust foundation for modelling feedback-driven processes through differential equations, accumulations (stocks), and transition rates (flows).

The structure of the proposed SD model draws inspiration from the project planning framework presented by Shafieezadeh et al. (2020), which simulates uncertainty and rework propagation in capital-intensive projects. That formulation incorporated structured work phases, deterministic review stages, and feedback-driven re-execution loops to represent project execution. The present model adapts these elements to R&D settings, extending them through direct interaction with an agent-based layer. Engineering management structures are simplified to reflect the inherent uncertainty and exploratory character of R&D, while deterministic interactions involving resources and infrastructure are maintained and captured through the ABM component, described in the next section.

Figure 1 presents the simplified stock-and-flow architecture of the SD model proposed in this work for R&D. Task effort progresses through four main stages—work to do, internal quality management, client review, and approved work. Each transition is governed by a flow equation that defines the rate at which effort moves from one stock to the next. Re-execution loops return part of the effort to previous stages when quality deficiencies or rejections occur, capturing the iterative feedback process characteristic of R&D activities.

The SD submodel does not represent individual events or decisions but instead captures the evolution of aggregate quantities, such as available tasks, completed work, and review workloads, over time. Differential equations define the rates of change (flows) between accumulated state variables (stocks), enabling the simulation of workload progression and rework cycles consistent with real-world project operations. Within the hybrid modelling framework, the SD model serves two primary functions: (i) aggregating micro-level outputs from the ABM into system-level indicators, and (ii) providing a dynamic representation of overall project behaviour to identify bottlenecks, rework dynamics, and schedule risks.

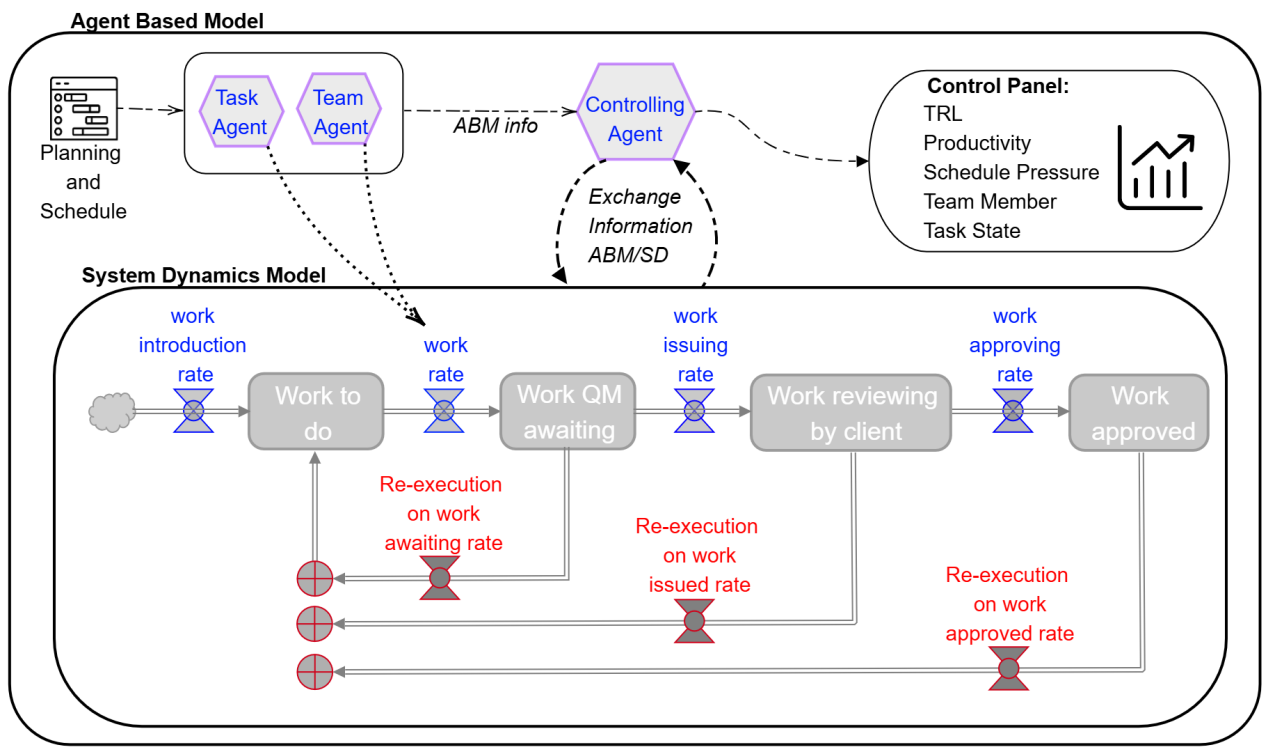

**Figure 1:** Simplified stock-and-flow diagram of the SD model. Rework is returned from multiple quality gates to the Work to do stock via separate re-execution flows.

### 2.1.3. Stock and Flow Structure

The SD submodel proposed in this work represents R&D project execution as a continuous flow of work evolving through successive development stages. The underlying architecture, shown in Figure 1, consists of interconnected stocks, flows, converters, and constants that describe the dynamic balance between completed, ongoing, and pending research activities.

| Stock | Description |
| --- | --- |
| $WTD_i$ | Work to do: accumulated R&D activities or experiments planned but not yet initiated, representing the project's backlog of scientific or engineering tasks. |
| $WQA_i$ | Work under quality assessment: ongoing research activities undergoing internal validation, data analysis, or prototype testing within the team. |
| $WCR_i$ | Work under client review: deliverables or reports under external evaluation by funding agencies, partners, or industrial stakeholders. |
| $WA_i$ | Work approved: completed and validated work that has successfully passed all internal and external reviews, contributing to project maturity and TRL progression. |

**Table 1**
Stock variables in the SD model, representing cumulative work at key R&D stages. Each stock is indexed by task $i$, preserving task-level resolution across the project execution. All variables are expressed in units of work (e.g., person-weeks).

In the context of RD, stocks correspond to cumulative states of work or knowledge at specific stages of the project. As summarised in Table 1, four main stocks define the system:

- Work to do ($WTD_i$): represents planned research activities or experimental tasks awaiting execution, analogous to the project's backlog of scientific or engineering work.

- Work under quality assessment ($WQA_i$): denotes ongoing verification and validation processes, such as data analysis, prototype testing, or internal peer review, which determine whether intermediate outputs meet technical requirements.

- Work under client review ($WCR_i$): captures the stage where deliverables are externally assessed by stakeholders or funding agencies, equivalent to external evaluations, audits, or collaborative review meetings in R&D programmes.

- Work approved ($WA_i$): represents completed and validated work that has successfully passed all internal and external assessments, contributing to technological progress and readiness.

Transitions between these stages are regulated through flows, listed in Table 2. Forward flows, such as the work introduction rate ($WIR_i$) and work rate ($WR_i$), control the initiation and progression of research activities. In contrast, re-execution flows ($RE_i$) return a fraction of the workload to earlier stages when new findings, experimental inconsistencies, or approval rejections require additional work. This cyclical mechanism captures the iterative learning process typical of R&D, where uncertainty, discovery, and refinement occur simultaneously.

Complementing these elements, Table 3 defines the auxiliary variables (converters) and constants that establish system-wide indicators. Converters include variables such as total work in process ($WIP_i$), which aggregates ongoing effort across all stocks, and realised project duration ($D_{act}$), which reflects the effective time required for project completion. Constants such as baseline duration ($D_{base}$) and duration scaling factor ($\gamma$) define reference values and scaling relationships that modulate project timing under varying conditions of resource availability or complexity.

All variables are indexed by task ($i$), preserving task-level heterogeneity throughout the model. This indexing allows detailed tracking of progress across different R&D activities, ensuring that variations in scope, complexity, and rework intensity are explicitly represented. The units and valid domains of all variables are provided in Tables 1, 2, and 3.

The SD structure illustrated in Figure 1 is proposed here to be mathematically represented through the following system of differential equations, formulated according to the stock–flow relationship defined in equation 2.

$$\frac{dWTD_i}{dt} = WIR_i - WR_i + RE_{QM,i} + RE_{c,i} + RE_{a,i} \tag{3}$$

| Flow | Description |
|---|---|
| $WIR_i$ | Work introduction rate: inflow of new R&D tasks generated from project planning or agent-based inputs. |
| $WR_i$ | Work rate: rate at which tasks progress from execution to internal quality assessment, influenced by team performance and workload. |
| $WIR_{QM,i}$ | Quality management release rate: transition flow from internal validation to external review, reflecting deliverables ready for client evaluation. |
| $WAR_i$ | Work approval rate: flow of tasks that complete all review stages and become finalised outputs contributing to TRL advancement. |
| $RE_{QM,i}$ | Re-execution rate (quality management): portion of work returning to execution due to failed internal validation or test results. |
| $RE_{C,i}$ | Re-execution rate (client review): work returned to previous stages after external rejection or requests for revision. |
| $RE_{A,i}$ | Re-execution rate (approved work): minor rework or post-validation adjustments prompted by new findings or process refinements. |

**Table 2**
Flow variables in the SD model, representing rates of transition between R&D stages. Each flow is indexed by task $i$, preserving task-level resolution across the project structure. All variables are expressed in units of work per week.

| Converter | Description | Unit | Domain |
|---|---|---|---|
| $WIP_i$ | Work in process: total active R&D effort aggregated across all project stages (sum of $WTD_i$, $WQA_i$, and $WCR_i$). Represents the total workload currently in execution or under review. | Work units | $[0, \infty)$ |
| $D_{act}$ | Actual project duration: cumulative time elapsed from the start of R&D activities to the completion of the last approved deliverable. Reflects the realised duration of the project. | Weeks | $[0, \infty)$ |
| **Constant** | **Description** | **Unit** | **Domain** |
| $D_{base}$ | Baseline duration: planned or reference duration for project execution, used as a benchmark to compare realised performance. | Weeks | Fixed |
| $\gamma$ | Duration scaling factor: proportionality constant that converts total effort into time, accounting for productivity variations or resource allocation levels. | Work units/week | Fixed |

**Table 3**
Converters and constants in the SD model, linking accumulated work and temporal performance in R&D projects.

$$\frac{dWQA_i}{dt} = WR_i - WIR_{QM,i} - RE_{QM,i} \quad (4)$$

$$\frac{dWCR_i}{dt} = WIR_{QM,i} - WAR_i - RE_{c,i} \quad (5)$$

$$\frac{dWA_i}{dt} = WAR_i - RE_{a,i}, \quad (6)$$

From a systems perspective, the differential equations describe the dynamic balance of effort as it moves through successive R&D stages. The first equation governs the inflow of new tasks and the redistribution of effort due to rework, capturing how unresolved issues feed back into the initial workload. The second and third equations represent internal and external validation processes, where effort transitions between quality assessment and client review while accounting for possible feedback loops arising from failed evaluations. The final equation formalises the accumulation of approved work, linking successful validation to the overall technological progress of the project. Together, these coupled equations define a structure in which work introduction, progression, validation, and rework continuously interact. This formulation allows the system to reproduce emergent behaviours such as learning cycles, schedule delays, and the nonlinear relationship between effort allocation and project maturity—key characteristics of real-world R&D processes.

## 2.2. Hybrid Agent-Based Modelling for R&D

The proposed SD modelling approach, Section 2.1.1, provides a comprehensive view of how R&D projects evolve through aggregated feedback and flow mechanisms. Nevertheless, the aggregate nature of SD limits its ability to capture heterogeneity in individual tasks and decentralised decision-making processes that define real R&D environments (Fu and Xing, 2021). In these settings, project outcomes depend on the interactions among diverse actors—researchers, engineers, and managers—whose decisions on resource use, scheduling, and problem-solving collectively shape system performance.

In our proposed framework, Agent-Based Modelling (ABM) complements SD through its capacity to represent such decentralised, micro-level interactions. In ABM, systems are described as populations of autonomous agents that operate under behavioural rules and interact within a shared environment. These local interactions give rise to emergent patterns that reflect the collective behaviour of the system (Helbing, 2012; Crooks et al., 2017). Within R&D contexts, agents can represent individuals or teams who execute research tasks, exchange information, and adapt their behaviour in response to evolving project conditions such as funding availability, uncertainty, and technological barriers.

ABM has been widely used in innovation and policy studies to explore adaptive behaviour, distributed learning, and coordination. Schlaile et al. (2020) developed a model to simulate knowledge diffusion in innovation networks, where agents assimilated new knowledge according to its compatibility with their existing expertise, reflecting absorptive capacity and cognitive diversity. Vinyals et al. (2023) employed ABM combined with reinforcement learning to analyse institutional strategies for innovation diffusion in agriculture, illustrating how behavioural diversity and policy experimentation shape adoption patterns. Similarly, Lopolito et al. (2013) examined the creation of innovation niches, showing how incentives such as subsidies and information sharing influence networking and knowledge exchange in emerging technological domains. These studies demonstrate how ABM enables the investigation of decentralised adaptation, learning, and coordination processes that are difficult to represent with aggregate-level methods such as SD.

The integration of ABM with SD combines the strengths of both paradigms, linking continuous system-level feedback with discrete, agent-level interactions. SD effectively models aggregated dynamics over time, yet it cannot explicitly represent task-level dependencies, sequential processes, or heterogeneous decision-making (Sterman, 2002). ABM, conversely, captures these aspects through decentralised, rule-based interactions but lacks the formal mathematical representation of continuous feedback that SD provides (Crooks et al., 2017). Merging these approaches enables the simultaneous simulation of micro-level execution dynamics and macro-level performance within a unified framework.

Previous studies have demonstrated the potential of hybrid AB–SD frameworks in diverse domains such as technology diffusion, health policy, and environmental systems (Swinerd and McNaught, 2014; Langarudi et al., 2021; Calderon-Tellez et al., 2024). However, applications to R&D project management sectors—remain scarce. The literature shows that most efforts either focus on SD-based feedback structures or on ABM-based agent interactions, with limited attempts to integrate the two (Shafieezadeh et al., 2020; Fu and Xing, 2021; Lopolito et al., 2013). This gap restricts the ability to capture the multi-level nature of R&D processes, where micro-level behaviours and macro-level feedback mechanisms jointly determine long-term technological progress and project outcomes.

The hybrid model proposed in this work addresses this gap through a combined AB–SD approach designed for R&D projects. The model simulates task-level and project-level feedback dynamics within a single simulation environment, with a particular focus on electrification and low-carbon innovation. It links the agent-level representation of individual or team behaviour with the system-level feedback structures that describe cumulative project evolution. The architecture of the model is illustrated in Figure 1, which shows the exchange of information between system variables in the SD layer and agent behaviours in the ABM component.

This hybrid framework serves as a proof-of-concept platform for exploring R&D execution dynamics in complex industrial ecosystems such as offshore electrification and energy transition technologies. It enables systematic analysis of how micro-level mechanisms—task heterogeneity, learning-driven productivity, and iterative rework—collectively shape macro-level outcomes, including schedule performance, resource utilisation, and technological maturity under uncertainty and interdependence.

## 2.3. Architecture and Coupling Logic

The previous subsections presented the SD and ABM components of the proposed hybrid framework, outlining their respective roles in representing the macro-level feedback structures and the micro-level behavioural dynamics of R&D projects. This subsection introduces the formal mathematical architecture that enables their integration. The objective is to define how information flows between the agent layer and the system-level model, establishing a coherent coupling logic that allows the microstate of individual agents to influence the aggregate variables of the SD component. The framework achieves this integration through an aggregation function that transforms the collective behaviour of task agents into system-wide indicators, which are then used to update the SD equations governing workload, progress, and rework dynamics. Conversely, outputs from the SD layer inform the evolving states and decisions of the agents, creating a feedback loop between local actions and global system evolution. The following formulation describes this architecture and the underlying coupling mechanism that supports synchronisation between discrete agent interactions and continuous system dynamics.

Let $A$ denote the set of agents in the system, where each agent $a \in A$ is characterized by a state $x_a(t)$ in the overall state space $X$ at time $t$. The collection of all agent states forms the system's microstate, $X_t$, and can be represented as,

$$X_t = \{x_a(t) : a \in A\} \subseteq X. \tag{7}$$

An aggregation function is used to extract system-level indicators from the microstates to allow the ABM to influence the SD model. In order to couple the agent layer with the SD model, relevant properties from the *Task* agent are aggregated into a set of macro-indicators, denoted as $S_t \in \mathbb{R}^n$. This aggregation is computed at each simulation step by the hybrid simulaton environment, SPM, described in Section 2.5. The *Controlling* agent operating within the ABM layer, accesses these agent-level values and exposes them as global properties for intra-agent coordination. Specifically, it accumulates the remaining effort across all task agents in selected states. Formally, the aggregation function $V : \mathcal{P}(X) \to \mathbb{R}^n$ is defined as,

$$S_t = V(X_t) = \left(\sum_{a \in A} w_i(x_a(t))\right)_{i=1\ldots n}. \tag{8}$$

Here, $w_i(x_a(t))$ are weighting functions that extract task-specific contributions to each aggregated macro-indicator, $S_t$, based on the agent's current state, $x_a(t)$. In the hybrid model, these functions are used to compute the total remaining effort in some of the *Task* agent states,

$$
\begin{aligned}
w_1(x_a(t)) &= \begin{cases} \texttt{remaining\_effort}_a(t), & \text{if } x_a(t) = \texttt{"open"} \\ 0, & \text{otherwise} \end{cases} \\
w_2(x_a(t)) &= \begin{cases} \texttt{remaining\_effort}_a(t), & \text{if } x_a(t) = \texttt{"in\_progress"} \\ 0, & \text{otherwise} \end{cases} \\
w_3(x_a(t)) &= \begin{cases} \texttt{remaining\_effort}_a(t), & \text{if } x_a(t) = \texttt{"rework"} \\ 0, & \text{otherwise} \end{cases}
\end{aligned}
\tag{9}
$$

The resulting macro-indicators, $S_t$ are passed as exogenous inputs to the SD model at each timestep, where they influence the flow rate of the SD stocks. For instance, the `in.progress.effort` affects the work rates, $WR$, while `rework.effort` affects the re-execution rates, $RE$. These flow rates then govern the stock variables, $WTD$, $WQA$, $WCR$, which describe the system-level distribution of work in different phases.

In the opposite direction, the SD model produces system-wide performance metrics (e.g., `schedule.pressure`) based on the evolving state of its stock. These macro-indicators are exposed back to the ABM layer and can influence agent behavior. For example, increased `schedule.pressure` may lead to higher `productivity` or altered decision rules among *Team Member* agents. This feedback is operationalized through an update function,

$$U : \mathcal{P}(X) \times \mathbb{R}^n \to \mathcal{P}(X), \quad X_{t+1} = U(X_t, S_t). \tag{10}$$

Where $U$ is the rule that determines how agent states evolve from one timestep to the next, based on both the agent's prior state and the current system-wide indicators.

Together, this feedback structure defines the core simulation logic of the hybrid model,

$$X_{t+1} = U(X_t, V(X_t)). \tag{11}$$

This recursive formulation establishes the dynamic coupling between micro-level agent states and macro-level system dynamics. The model represents how individual task interactions collectively determine project-wide conditions and how these aggregated outcomes feed back to influence agent decisions in subsequent iterations. This bidirectional exchange forms a co-evolutionary process in which local and global behaviours continuously shape one another. Through this structure, the hybrid framework enables a systematic examination of interdependencies and feedback mechanisms that characterise R&D project environments. It also provides a means to analyse how coordination strategies, workload propagation, and schedule pressure emerge from agent interactions and evolve under changing execution conditions, offering a foundation for exploring policy and management interventions in complex innovation systems.

### 2.4. Relating Project Execution to Cumulative Distribution of Technology Redness Level

The hybrid AB–SD framework presented in the previous sections provides a dynamic representation of how R&D projects evolve through interconnected feedback and task-level interactions. While this structure captures the mechanisms driving project execution, it remains necessary to link these operational dynamics to a measurable indicator of technological maturity. This section introduces a probabilistic formulation that connects simulated project progress to TRLs, thereby extending the model's capability to assess the likelihood that a technology will attain specific maturity thresholds over time.

Although the proposed hybrid model does not directly rely on empirical TRL milestone data, it enables the estimation of technological maturity through a statistical representation of cumulative project progress. The formulation establishes a probabilistic relationship between task completion and maturity attainment, allowing the interpretation of simulated project outcomes within a TRL-oriented framework.

Let $N(t)$ denote the cumulative number of completed tasks at time $t$, and let $N_{\text{total}}$ represent the total number of tasks within the project. A normalized measure of progress can then be defined as:

$$x(t) = \frac{N(t)}{N_{\text{total}}} \in [0, 1] \tag{12}$$

which serves as a proxy for technological advancement under the assumption that each completed task contributes incrementally to the overall system maturity.

To represent the uncertainty and variability inherent in R&D execution, a cumulative distribution function (CDF) is derived from the logarithmic transformation of $x(t)$. The distribution of $\ln(x)$ is modelled through the Student's t-distribution, which provides flexibility for limited sample sizes and captures the stochastic nature of progress trajectories. Following the approach of Kenley et al. (2024), the cumulative probability of achieving a given maturity level is expressed as,

$$F(x) = P(\text{Maturity}|x) = t_\nu\left(\frac{ln(x) - \hat{\mu}}{\sigma}\right) \tag{13}$$

where $t_\nu(\cdot)$ denotes the cumulative distribution function of the Student's t-distribution with $\nu$ degrees of freedom. The parameters $\hat{\mu}$, $S$, and $\nu$ are estimated from multiple simulation runs $x_k^*$, each corresponding to a realization of normalized progress, and they are defined as follows,

$$\hat{\mu} = \frac{1}{n}\sum_{k=1}^{n} \ln(x_k^*), \tag{14}$$

$$\sigma = \sqrt{\frac{1}{n-1}\sum_{k=1}^{n}\left(\ln(x_k^*) - \hat{\mu}\right)^2}, \tag{15}$$

$$\nu = n - 1. \tag{16}$$

Aggregating results from multiple stochastic simulation runs produces an empirical maturity profile that evolves with time. This formulation does not replace formal TRL assessments but offers a quantitative approximation of the probability that a given technology achieves a specific readiness level, conditional on simulated execution dynamics. From a systems perspective, this probabilistic mapping bridges the operational layer of project execution with the strategic layer of technology assessment. It embeds TRL evolution within the hybrid simulation framework, transforming execution-based progress metrics into interpretable maturity indicators. The resulting maturity distribution supports decision-making under uncertainty, enabling project managers and policymakers to evaluate development trajectories, identify bottlenecks in technological advancement, and prioritise actions that accelerate readiness. In this way, the integration of TRL modelling completes the methodological framework introduced in this paper, linking task-level behaviour, system-level feedback, and maturity evolution into a unified analytical approach for R&D management.

### 2.5. Integration Engine the Role of the SPM

The hybrid AB–SD framework introduced in the previous section requires a coordinated mechanism to manage the interaction between its discrete and continuous components. This coordination is achieved through the Simulation Process Manager (SPM), which acts as the central integration engine of the model. The SPM governs the execution of simulation steps, synchronises data exchange between the ABM and SD layers, and ensures temporal consistency throughout the simulation process.

Within the agent-based layer, the SPM instantiates and manages the lifecycle of the main agent classes—*Task*, *Team Member*, and *Controlling* agents. These agents collectively represent the decentralised decision-making and behavioural adaptation characteristic of R&D project execution, encompassing task progression, resource allocation, learning effects, and responses to schedule conditions. The SPM evaluates agent interactions and state transitions at each simulation step, maintaining coherence among interdependent activities.

Simultaneously, the SPM interfaces with the continuous-time SD model, which tracks aggregated project-level indicators such as total work in progress, quality assessment flows, and rework dynamics. Information generated within the ABM—such as the cumulative remaining effort, task completion rate, and rework demand—is aggregated and transferred to the SD layer as exogenous variables. The SD component, in turn, calculates system-wide indicators such as `schedule.pressure` and exposes them to the ABM, where they influence agent decision rules, productivity levels, and task prioritisation.

The SPM therefore ensures a consistent feedback exchange between both modelling layers. Each simulation cycle follows a structured sequence: (i) agents update their states according to local conditions, (ii) aggregated variables are passed to the SD model, (iii) system-level feedback is computed, and (iv) the resulting indicators modify subsequent agent behaviour. This bidirectional integration maintains logical consistency between micro- and macro-level dynamics, allowing the exploration of multi-scale dependencies and emergent coordination patterns within R&D projects.

*Behavioural and Structural Components of the Hybrid Model*

Schedule pressure is a central construct in the hybrid model, representing the urgency to complete remaining work given the constraints of time and team capacity. It is formalized as a continuous, dimensionless index calculated as the ratio of total remaining effort, $E_{\text{remaining}}$, to the product of remaining project time, $t_{\text{remaining}}$ and available team members, $N_{\text{team}}$,

$$\text{SP} = \min\left(\frac{E_{\text{remaining}}}{t_{\text{remaining}} \cdot N_{\text{team}}}, 3.0\right) \tag{17}$$

This formulation ensures that schedule pressure increases as the deadline approaches or as the workload accumulates, but is capped at a maximum value of 3.0 to avoid unrealistic escalation in urgency. This assumption is consistent with established project dynamics models that associate pressure with cognitive and performance-related trade-offs (Sterman, 2002).

The effect of schedule pressure on productivity is implemented through a non-linear lookup function. The productivity increases with moderate pressure, but declines beyond a certain threshold, where excessive urgency leads to cognitive overload and error rates (Kuutila et al., 2020).

The resulting productivity-pressure feedback loop makes the model able to simulate how variations in workload and time availability influence execution dynamics. By dynamically adjusting the global productivity factor based on schedule pressure, the model captures how performance levels respond to perceived urgency. This mechanism allows for the emergence of reinforcing cycles, such as schedule-driven productivity gains that later collapse into rework or delays.

Team members are modeled as homogeneous agents. Each agent possesses identical properties in terms of availability, skill level, and behavioral logic for task selection and execution. At every simulation timestep, available agents evaluate the collection of eligible tasks, specifically those in the `open` or `rework` states that are unassigned and have all dependency constraints resolved. Task selection follows a simple heuristic: agents assign themselves to the first valid task encountered based on the scenario-defined ordering.

To account for individual performance evolution over time, the model incorporates an endogenous learning mechanism. Each team member's productivity increases as a function of experience, operationalized through a sigmoid learning curve. Specifically, as an agent completes more tasks, their personal productivity rises rapidly at first, and then asymptotically approaches a predefined maximum or a saturation point. These learning dynamics are well-supported in innovation systems literature as an influent aspect to the project performance (Hekkert et al., 2007; Beste et al., 2020).

In addition to the individual learning trajectory, team productivity is also modulated by the system-wide schedule pressure index, as described earlier. This dual feedback structure, combining local experience-based learning with global schedule-driven adaptation, enables the model to explore non-linear productivity trajectories throughout the execution of the project. It captures realistic behavioral shifts commonly observed in innovation-intensive project environments, where individuals both learn from experience and adapt to systemic urgency. While this abstraction also neglects certain domain-specific learning rates or specialization effects, it preserves core behavioral feedbacks that shape R&D execution under uncertainty and time pressure. Comparable modeling choices are found in dynamic simulation studies of project resource management, such as Lin and Dai (2014) work on Petri net modeling of workforce dynamics.

*Task Dynamics and Structural Dependencies*

In the proposed hybrid AB–SD model, each R&D task is represented as an autonomous agent with fixed properties such as initial effort, rework probability, and predefined review delays. The initial attempt assigned to a task refers to the total amount of work required to complete it, expressed in abstract work units that capture the relative complexity, uncertainty, and resource intensity of each development activity, rather than real labour hours. These scenario-defined values draw from established concepts in staged technological development, including the TTRL framework and the S-curve theory of innovation progression (Lavoie and Daim, 2017; Buchner et al., 2019; Lezama-Nicolás et al., 2018). Within this formulation, early-stage tasks are assumed to demand higher effort due to exploratory research, conceptual design, or experimentation, while later-stage tasks focus on refinement, integration, and validation. This reflects the common pattern in RD processes, where uncertainty and learning intensity decline as the project moves toward maturity.

Task interdependencies are defined prior to simulation and remain static throughout execution. A task can only start once all of its designated predecessor tasks have been completed, establishing a directed dependency network. This structure allows for the controlled examination of systemic effects such as bottleneck formation, congestion in dependency chains, and the propagation of rework or delays through sequential phases. Although real-world R&D projects frequently exhibit evolving task relationships in response to discoveries or external constraints, the use of static dependencies simplifies analysis while preserving the essential causal feedbacks between interdependent development activities. This design enables the exploration of how structural properties—such as dependency depth or task clustering—influence overall project performance and technological advancement.

Each task progresses through a defined sequence of states that represent the core phases of the R&D execution cycle, as detailed in Section B.1. After a task is initiated by a *Team Member* agent and executed to completion, it enters two successive validation stages: an internal quality management (QM) review and an external client review. These stages are modelled as deterministic delays with default durations of 1.0 and 2.5 weeks, respectively.

This representation mirrors the structured review processes observed in R&D project management, where technical validation and stakeholder feedback function as essential control points in the innovation cycle. The deterministic formulation excludes stochastic variation, queuing effects, and resource constraints, which are outside the scope of this initial model but can be incorporated in future refinements.

This modelling choice is supported by the work of Shafieezadeh et al. (2020), who demonstrated that explicit quality control stages in simulation models are essential for reproducing rework dynamics, performance feedback, and decision-making under uncertainty. In a similar manner, the review phases in the proposed model act as analytical checkpoints through which the effects of delay, rework, and information feedback can be observed. They enable investigation of how errors detected during validation influence schedule performance and workload redistribution across the project. While simplified, this formulation maintains interpretability and analytical transparency, providing a foundation for future extensions that incorporate stochastic review outcomes, concurrent verification processes, or variable resource capacities.

Tasks are modelled as heterogeneous, interdependent agents linked through structured review mechanisms. This formulation captures the iterative and feedback-driven nature of R&D execution and connects micro-level task behaviour to the macro-level feedback processes simulated within the System Dynamics layer. The resulting integration supports investigation of how completion patterns, validation loops, and structural dependencies collectively shape project performance and the trajectory of technological maturity.

Finally, rework, in the hybrid AB-SD model, represents the need for task revision following inadequate outputs during internal or external review phases. At the agent level, rework is triggered probabilistically after a task has completed client review. Each *Task* agent is initialized with a predefined rework probability, typically ranging from 0.2 to 0.5. These probabilistic values determine the likelihood of being returned for re-execution. If triggered, the task transitions into a `rework` state, and a portion of its original effort is reassigned for completion. This abstraction is based on observed patterns in R&D and engineering projects, where tasks in earlier development stages are frequently more subject to adjustment due to technical immaturity than tasks in later stages. Literature on this demonstrates that structured review checkpoints and re-execution phases are critical mechanisms for managing uncertainty and detecting performance gaps (Shafieezadeh et al., 2020; Calderon-Tellez et al., 2024).

At the system dynamics layer, the cumulative remaining effort of all *Task* agents in the `rework` stage is aggregated by the *Controlling* agent and passed to the SD model as the variable `abm.rework.effort`. This aggregated indicator governs the re-execution flows within the SD model, as detailed in Section 2.3. The current implementation assumes that high-level effort measures sufficiently capture the magnitude of ongoing rework activity.

*Parameterization and Scenario Inputs*

All simulation experiments are executed under predefined scenario configurations contained in a structured input file. This configuration defines fixed values for core parameters, including task attributes, agent decision rules, and system-wide constants. These parameters establish the boundary between the conceptual model and the simulation environment, ensuring consistency and reproducibility across experimental runs. The external configuration approach supports transparent hypothesis testing and comparative analysis while avoiding the inclusion of context-specific assumptions within the model structure itself. Scenario parameters can be adjusted without modifying the source code, facilitating sensitivity studies and systematic exploration of different RD settings.

The integration of the ABM and SD layers represents the central feature of the hybrid framework. The coupling design follows the conceptual structure outlined by Howick et al. (2024) and is implemented through bidirectional data exchange summarised in Table 4. At each simulation step, agent-level task states and remaining efforts are aggregated into system-wide indicators that update the SD stocks and flows. In return, the SD layer produces global feedback variables, such as `schedule.pressure`, that influence agent productivity and task selection behaviour. This reciprocal exchange maintains coherence between micro-level interactions and macro-level dynamics, allowing emergent behaviours to evolve consistently within the simulation.

The model operates as a proof-of-concept framework addressing a critical methodological gap in RD modelling. Its objective is to demonstrate how a hybrid AB–SD approach can capture both execution dynamics and the probabilistic progression of technological maturity. The focus, therefore, lies in conceptual exploration and theoretical grounding. This approach aligns with the established role of system simulation as an exploratory method for studying complex socio-technical systems (Sargent, 2010).

Validation of the model focuses on three complementary dimensions: structural, behavioural, and hybrid consistency. Structural validation assesses whether the model's internal logic aligns with the theoretical understanding of RD

| Interface | Description |
| --- | --- |
| *Information flows from SD model to ABM model* | |
| (1) Stock levels influence agent parameters | The SD model computes global project-level variables such as schedule pressure, which are exposed to all agents as external inputs. These values influence agent decision logic, such as how quickly assigned team members complete work. |
| (2) Feedback propagation to agent performance | Fluctuations in SD stocks reflecting system-wide workload indirectly affect agent efficiency by modifying the shared productivity, simulating systemic schedule strain. |
| *Information flows from ABM model to SD model* | |
| (3) Aggregated task state effort drives SD flows | The SD model consumes agent-level task states and remaining effort to compute aggregate values such as abm.in.progress.effort, abm.open.effort, and abm.rework.effort. These aggregated measures define the inflows and outflows in the SD stock-flow structure. |
| (4) Behaviors of agents affect flows | Behaviors of agents in an ABM model can influence flows in an SD model by increasing/decreasing parameters used in equations for flows. |
| (5) Agent task completions influence maturity dynamics | The cumulative number of closed task, tracked within the agent layer, informs the SD-level computation of technological maturity. This allows the model to translate micro-level execution into macro-level readiness assessments. |

**Table 4**
Biodirectional integration interfaces between ABM and SD layers in the hybrid model projects. The consistency of the SD and ABM layers was verified through logic checks and parameter sensitivity tests to confirm expected directional behaviour under varying conditions. Behavioural validation assesses whether simulation outputs reproduce recognised patterns of RD system behaviour, such as S-curve growth in technological maturity, delay propagation in sequential task chains, and performance constraints under limited resources. Hybrid-specific validation evaluates the correctness of interactions between the ABM and SD components, ensuring accurate data exchange and alignment of aggregated measures used for TRL estimation (Li et al., 2024).

Several factors constrained large-scale empirical verification, including the proprietary nature of detailed RD project data, the non-stationary characteristics of innovation systems, and the limited access to industry experts within the current research scope. Despite these challenges, the model's structure enables case-specific validation through expert judgement and historical comparison. This approach allows alignment of simulated patterns with expert understanding of RD project dynamics and reference to documented historical cases, producing an expert-level empirical assessment of the model's behaviour. Such validation, while qualitative in nature, ensures that the simulation outcomes remain grounded in real-world experience and established knowledge. The results presented in the following section were obtained in order to confirm whether the hybrid AB–SD framework provides a consistent, transparent, and extensible foundation for continued refinement and domain-specific applications.

## 2.6. Model Hypotheses and Future Development Directions

The hybrid AB–SD framework presented in this work represents an initial step toward a formal and integrated modelling approach for analysing the dynamics of R&D projects. Its purpose is to establish a foundational structure that connects individual task execution, team interactions, and system-level feedback mechanisms to technology maturity progression. While the model provides a consistent analytical foundation, its formulation incorporates a set of simplifying hypotheses designed to make the problem tractable and to prioritise conceptual clarity over empirical specificity. These hypotheses are summarised below, together with the reasoning that supports each choice.

- **Homogeneous task contribution:** Each task contributes proportionally to the cumulative technological maturity of the project. The proportion of completed tasks is treated as a proxy for readiness progression, assuming that all tasks have an equivalent marginal effect on the overall TRL evolution. *Rationale:* This simplification supports the first formalisation of the coupling between task execution and TRL prediction. It preserves transparency and interpretability in the model's initial version. Future developments can incorporate heterogeneous maturity weights based on task complexity, novelty, or criticality within the technological system.

- **Simplified resource allocation:** Team resources are represented through a continuous variable that modulates overall productivity. *Rationale:* This abstraction keeps the analysis centred on behavioural adaptation and system feedbacks, avoiding the computational burden of discrete scheduling or optimisation algorithms. Later versions of the model can integrate adaptive allocation mechanisms or reinforcement learning strategies to simulate resource prioritisation under uncertainty.

- **Representative agent behaviour:** Agents follow rule-based decision logic that reproduces typical behavioural patterns in R&D teams—such as productivity adjustment, learning progression, and rework initiation—without modelling individual personality or cognitive variation. *Rationale:* This approach enables exploration of collective learning and coordination effects while maintaining computational tractability. Subsequent refinements may include more detailed behavioural rules or cognitive decision-making processes derived from empirical data.

- **Aggregated feedback representation:** System-level indicators, including schedule pressure, workload accumulation, and resource utilisation, are computed as aggregate variables that apply uniformly across agents. *Rationale:* Aggregation ensures coherent information exchange between the ABM and SD layers, preserving the integrity of the coupling architecture. Future extensions could include network-based or hierarchical feedbacks to represent differentiated team structures or communication patterns.

- **Idealised project topology:** The R&D workflow follows a simplified linear progression through execution, internal review, client review, and approval stages, with feedback loops representing rework and learning. *Rationale:* This topology captures the cyclical and iterative essence of R&D without requiring project-specific workflow graphs. Future versions could incorporate parallel task streams, concurrent engineering structures, or dynamic task reconfiguration to capture the full complexity of industrial R&D projects.

These hypotheses define the scope of this first formulation and clarify that the model functions as an analytical tool for understanding R&D system behaviour rather than a predictive simulator for specific industrial cases. The approach establishes a scalable base that can be progressively refined with empirical data, domain-specific parameters, and advanced behavioural rules, as demonstrated in the practical implementation of this work, which will be presented in the results section.

Future research should focus on expanding three key dimensions of the framework: (1) introducing heterogeneous task characteristics and non-linear maturity relationships to better capture the uneven contributions of different R&D activities to technological progress, (2) integrating adaptive and learning-based decision mechanisms that enable agents to adjust behaviour dynamically in response to changing project conditions, and (3) extending validation through empirical calibration with data from a larger variety of active R&D programs, aimed at refining parameter estimates, enhancing behavioural consistency, and increasing the framework's predictive and explanatory depth.

Overall, this formulation represents a first systematic attempt to link agent-level execution, system-level feedback, and TRL progression within a unified hybrid model. It provides a conceptual and computational foundation that can be extended to address the complexity of modern R&D environments and to support strategic decision-making in emerging technological fields.

## 3. Results and Discussion

The following sections present the simulation experiment results obtained using the hybrid AB-SD model introduced in Section 2. The primary aim is to assess the model's internal behaviour, sensitivity, and structural plausibility under varying conditions relevant to R&D project management.

### 3.1. Base Case

Given the limited availability of empirical datasets for early-stage R&D projects, the base scenario is designed as a representative reference experiment to examine the structural dynamics of innovation processes under controlled and transparent assumptions. This baseline configuration operationalises the hybrid AB–SD model introduced in Section 2, decomposing the project into 15 discrete tasks distributed over a 156-week planning horizon. Each task is characterised by a specific workload and a probabilistic rework rate, capturing both technical uncertainty and iterative feedback processes that typically define high-uncertainty R&D environments.

In this configuration, tasks are executed in parallel without precedence constraints, forming a benchmark condition that isolates task-level variability and resource allocation effects from higher-order structural dependencies. This design serves as a reference model for subsequent comparative experiments, facilitating a systematic evaluation of how interdependencies, sequencing, and resource constraints impact system performance and technological progression.

| Task Number | Typical Tasks Covered | Effort Range [Work Units] | Rework Probability Range |
|---|---|---|---|
| 1–4 | Concept development, literature review, feasibility analysis | 16–22 | 0.42–0.50 |
| 5–8 | Prototype building, component validation | 13–16 | 0.28–0.45 |
| 9–11 | Demonstration activities, prototype testing | 10–12 | 0.15–0.25 |
| 12–15 | Integration, commercialization | 5–10 | 0.10–0.18 |

**Table 5**
Task group characteristics in the R&D simulation model

| Property | Value |
|---|---|
| deadline | 156 weeks |
| timestep | 1.0 weeks |
| team member | 5.0 |
| team.member.learning.factor | 0.2 |
| team.member.max.productivity | 3.0 |

**Table 6**
Reference values for model's parameters.

Task complexity and uncertainty are stratified across the stages of technological maturity, following the TRL framework for the chemical industry proposed by Buchner et al. (2018) (see Table 15). As detailed in Table 5, tasks are organized into four TRL-based phases, spanning from conceptual ideation (TRL 1–3) to system-level validation and pre-commercial integration (TRL 8–9). Early innovation stages (Tasks 1–4) encompass high-effort, high-risk activities such as concept formulation and feasibility testing, reflected in the largest workloads (16–22 work units) and rework probabilities (0.42–0.50). Intermediate stages (Tasks 5–11) represent prototyping and validation phases, characterised by progressive uncertainty reduction. Final stages (Tasks 12–15) correspond to late-stage development, with minimal rework risk (0.10–0.18) and lower task effort (5–10 work units), consistent with increasing technological maturity and stabilization.

The simulation operates at a temporal resolution of one week per timestep. Agent properties governing team member behavior are specified to reflect generalized R&D learning curves and bounded productivity profiles. As detailed in Table 6, each agent begins with baseline productivity, incrementally increasing performance over time through a learning coefficient of 0.2, subject to a maximum output ceiling of 3.0 work units per week. Empirical and theoretical studies inform these parameter values on performance dynamics in knowledge-intensive project environments.

### 3.1.1. Experiment 1 - Dependencies

The first experimental scenario investigates the structural implications of task interdependencies on project execution dynamics within the hybrid AB–SD simulation framework. Drawing upon the modelling logic and assumptions about interdependencies introduced in Sections 2.2 and 2.6, the experiments contrast two extreme configurations: (i) a parallel task structure, representing the base-case scenario, and (ii) a strictly sequential configuration, in which all fifteen project tasks are subject to linear precedence constraints.

In the parallel configuration, all tasks are assumed to be mutually independent, allowing simultaneous execution provided that agent availability is met. This abstraction represents an idealised condition of maximum concurrency and absence of dependency-induced delays. In contrast, the sequential configuration requires each task to await completion of its predecessor, representing a more constrained and risk-averse project logic often applied in high-dependency environments such as critical path scheduling. All other model parameters, including task attributes, agent behaviour, simulation horizon, and performance coefficients, remain unchanged between configurations, as listed in Tables 5 and 6. This design isolates the causal effect of dependency structure, following System Dynamics best practices for testing endogenous relationships (Sterman, 2002).

Table 7 summarises the resulting performance metrics. Under parallel execution, the system exhibits high resource concurrency, with an average of 1.2 tasks in progress and a peak queue length of 10 tasks. The average cycle time, defined as the elapsed time from task initiation to closure, reaches 47.9 weeks. Flow efficiency, reflecting the proportion of active work relative to total cycle time, is moderate at 48.1%. Resource utilisation (23.7%) and throughput (0.2 tasks per week) remain stable, and the project completes within 72 weeks, well ahead of the 156-week deadline, achieving 100% on-time delivery.

In contrast, the sequential configuration produces markedly different dynamics. The enforced dependency chain extends project makespan to 171 weeks, reducing throughput to 0.09 tasks per week. Despite this, the average cycle time per task decreases to 25.6 weeks, and flow efficiency improves significantly to 91.2%. Queue lengths and resource contention diminish, but this comes at the cost of systemic inflexibility and temporal bottlenecking. On-time delivery falls to 66.7%, with the remaining 33.3% of tasks completed beyond the 156-week target.

Further differences emerge in task state occupancy, as summarised in Table 8. The cumulative task weeks spent in rework states decrease by 88% in the sequential case due to reduced compounding of concurrent rework feedback. Marginal reductions are also observed in client review and quality management delays. These findings are consistent with empirical observations reported in previous System Dynamics studies of project governance and feedback propagation (Calderon-Téllez et al., 2023). The results confirm that dependency-driven structures can reduce variability and rework propagation, but at the expense of adaptability and overall throughput.

To evaluate how task dependency structures influence perceived technological maturity under uncertainty, an empirical CDF was derived from multiple stochastic simulation runs for both parallel and sequential task execution configurations. This probabilistic measure, visualised in Figure 3, reflects the cumulative likelihood that the system has achieved a given level of technological readiness by a specific point in time. The CDFs were computed by aggregating normalised progress trajectories from 100 runs and smoothing them to highlight general trends. In this context, normalised progress is defined as the cumulative fraction of completed tasks over the project horizon, which serves as a proxy for maturity. Although this proxy assumes task contribution to system maturity, as described in Section 2.4, it enables a consistent and interpretable metric for comparative analysis of task dependency structures.

Figure 2 presents the temporal evolution of task states for the parallel and sequential configurations in Experiment 1. The graphs show how the number of tasks in each state changes over time and how this dynamic reflects differences in workflow coupling, concurrency, and completion patterns. In the parallel configuration (top plot), the number of open tasks decreases rapidly within the first 40 weeks as most activities are launched simultaneously. The in-progress curve initially rises steeply, reaching a peak corresponding to the full utilisation of available resources, and then gradually declines as tasks are completed. The closed curve increases quickly and stabilises well before the 156-week deadline, showing that all tasks are finalised within approximately half of the project duration. This pattern characterises systems with high concurrency, short feedback loops, and greater exposure to rework caused by simultaneous development. Such oscillations between open and in-progress states replicate the empirical dynamics observed in flexible R&D programmes, where overlapping work packages accelerate output but generate coordination noise and iterative correction cycles (Calderon-Téllez et al., 2024).

In the sequential configuration (bottom plot), the progression is more gradual. Only a few tasks are active at any given time, while others remain in the waiting dependencies state until their predecessors are completed. The closed curve increases stepwise, reflecting the gate-by-gate advancement typical of tightly coupled project networks. Completion occurs approximately 200 weeks after the start, exceeding the nominal 156-week deadline. This outcome reproduces the characteristic behaviour of stage-gated industrial R&D systems, where dependency constraints mitigate rework and uncertainty but limit flexibility and increase overall duration.

The model therefore captures well-established empirical trade-offs documented in system dynamics and project-management research:

(i) Concurrency–rework relationship: high parallelism enhances early progress but amplifies iteration cycles.

(ii) Sequential stability: ordered execution improves predictability and quality assurance but extends project duration.

(iii) Learning and coordination feedbacks: the damped oscillations and plateauing of completion rates reproduce typical learning-curve and workload-saturation effects observed in complex engineering and R&D settings (Sterman, 2002).

| Metrics | Parallel | Sequential |
|---|---|---|
| Average cycle time [week] | 47.9 | 24.9 |
| Flow efficiency [%] | 48.1 | 91.2 |
| Peak Queue [tasks] | 10.0 | 2.0 |
| Average queue [tasks] | 1.5 | 0.1 |
| Peak tasks in progress [tasks] | 5.0 | 5.0 |
| Avg tasks in progress [tasks] | 1.4 | 0.9 |
| Utilization [%] | 23.7 | 17.0 |
| Makespan [weeks] | 88.0 | 208.0 |
| Throughput [tasks/week] | 0.2 | 0.1 |
| On-time [%] | 100.0 | 66.7 |
| Late [%] | 0.0 | 33.3 |

**Table 7**
Performance metrics comparison for parallel (no dependencies) and sequential (strict/with dependencies) task execution scenarios.

| Task State | Parallel [task-weeks] | Sequential [task-weeks] | Change [%] |
|---|---|---|---|
| Rework | 25.0 | 3.0 | −88 |
| Client Review | 80.0 | 72.0 | −10 |
| QM Review | 40.0 | 36.0 | −10 |

**Table 8**
Cumulative task-weeks in rework, client review, and quality management review states for parallel and sequential task execution scenarios. A task-week is one task occupying a given state for one week.

Overall, the simulation results demonstrate that the hybrid AB–SD framework accurately reproduces the temporal and structural patterns observed in empirical project studies. The evolution of task states, the timing of saturation, and the trade-offs between flexibility and control are consistent with observed dynamics in real R&D environments, confirming the model's validity as a representation of technology-development workflows.

Figure 3 presents the cumulative distribution of technological maturity for the parallel and sequential configurations in Experiment 1. The parallel configuration exhibits a steeply rising maturity CDF, reaching a probability of approximately 90% by week 70 and plateauing near 1.0 around week 90. This pattern reflects the high throughput and efficiency reported in Table 7 and demonstrates the advantages of full task concurrency. The steep gradient of the curve indicates limited inter-run variability, consistent with low structural coupling and high schedule flexibility between tasks. The characteristic S-shape of the curve closely resembles classical logistic growth patterns used to describe Technology Readiness Level (TRL) progression in empirical studies (Lavoie Daim, 2017; Buchner et al., 2019). This similarity reinforces the conceptual validity of linking simulated execution performance with probabilistic maturity assessments, even in the absence of explicit TRL milestone data.

In contrast, the sequential configuration produces a slower and more extended maturity trajectory. The slope of the CDF is noticeably shallower, with the 90 % probability threshold not reached within the 156-week deadline and instead attained only after week 200. This delayed progression reflects the restrictive effect of strict precedence constraints, where each task must await the completion of its predecessor before initiation. The resulting dependency propagation introduces structural rigidity and greater temporal sensitivity, amplifying variability in cumulative completion across simulation runs. Consequently, the maturity CDF remains substantially lower within the planned timeframe, signalling an elevated risk of underperformance and schedule overrun.

Overall, the comparison between the two curves illustrates how the project structure directly affects the dynamics of technological maturity. Parallel configurations accelerate early-stage learning and convergence toward complete maturity. In contrast, sequential configurations ensure control and quality but delay systemic progress—a trade-off widely recognised in the literature on System Dynamics and RD project management (Sterman, 2002; Calderon-Téllez et al., 2024).

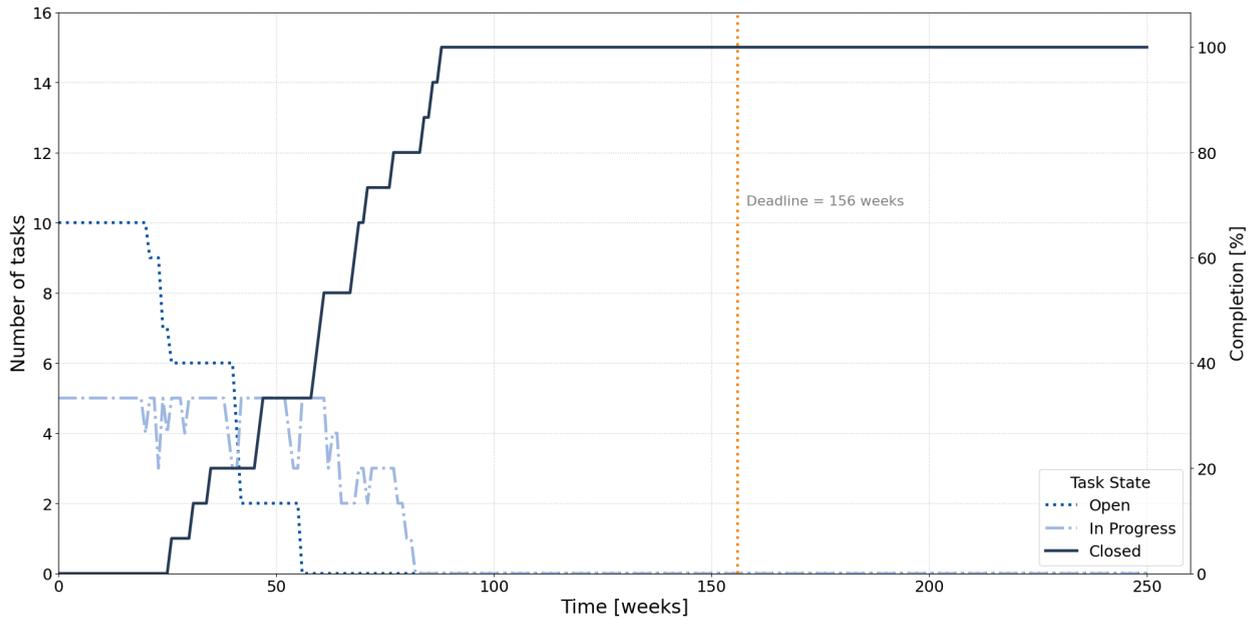

(a) Parallel

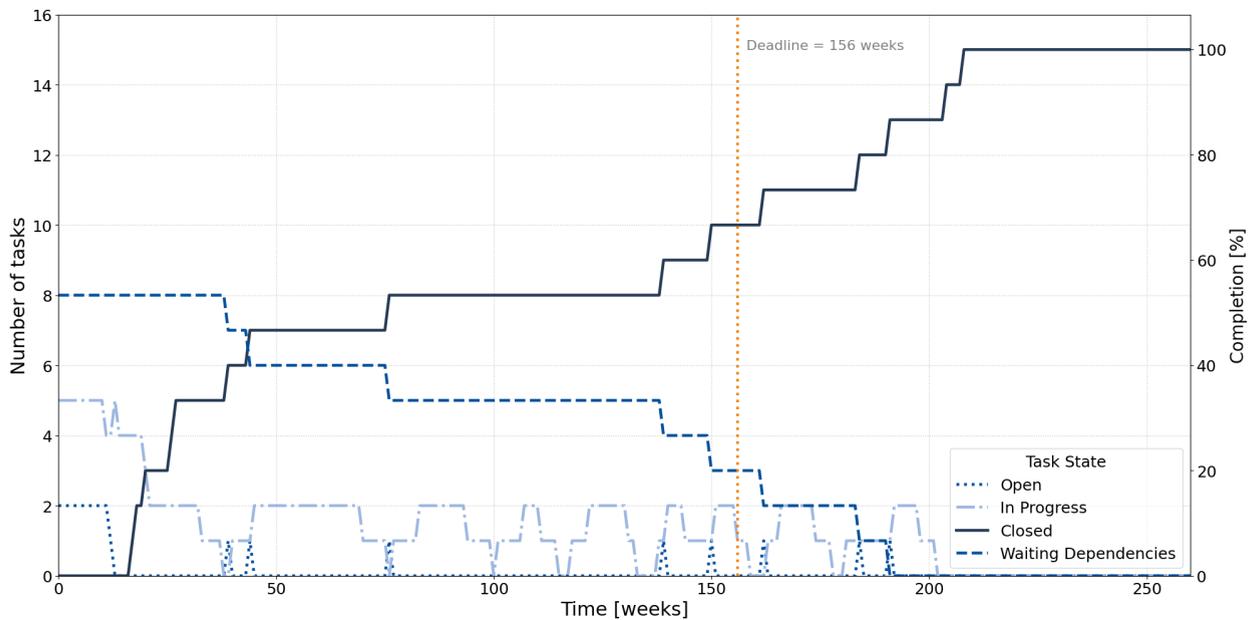

(b) Sequential

**Figure 2:**

The simulation results from Experiment 1 provide strong evidence that the proposed model successfully reproduces how task-dependency structures fundamentally influence both operational dynamics and emergent maturity trajectories in R&D projects. While the performance indicators reported in Table 7 clearly distinguish the outcomes of parallel and sequential configurations, these results must be interpreted from a systemic perspective to reveal their broader implications for project behaviour under uncertainty. From a system-dynamics standpoint, the emergence of consistent patterns across both task states and maturity CDFs validates the structural coherence of the hybrid AB–SD model. The parallel configuration demonstrates behaviour characteristic of loosely coupled systems, marked by high concurrency, low systemic friction, and greater freedom of operation. These conditions yield a shorter makespan and higher throughput, accompanied by a sharply rising cumulative maturity trajectory that follows the classical S-curve pattern widely recognised in TRL and innovation management literature. The smooth, sigmoidal increase of the maturity CDF indicates that in unconstrained configurations, the stochasticity of individual task execution aggregates into a predictable, self-reinforcing progression of technological maturity. This outcome aligns with established System Dynamics theory, which posits that feedback structures with minimal coupling tend to stabilise over time through reinforcing resource utilisation loops and adaptive learning effects. In this case, feedback on completed work accelerates subsequent progress by freeing up team capacity, reinforcing positive productivity–learning cycles, and sustaining high project momentum. Such behaviour mirrors empirical observations in real R&D environments, where parallel experimentation and rapid iteration often lead to non-linear yet stable growth in technological readiness.

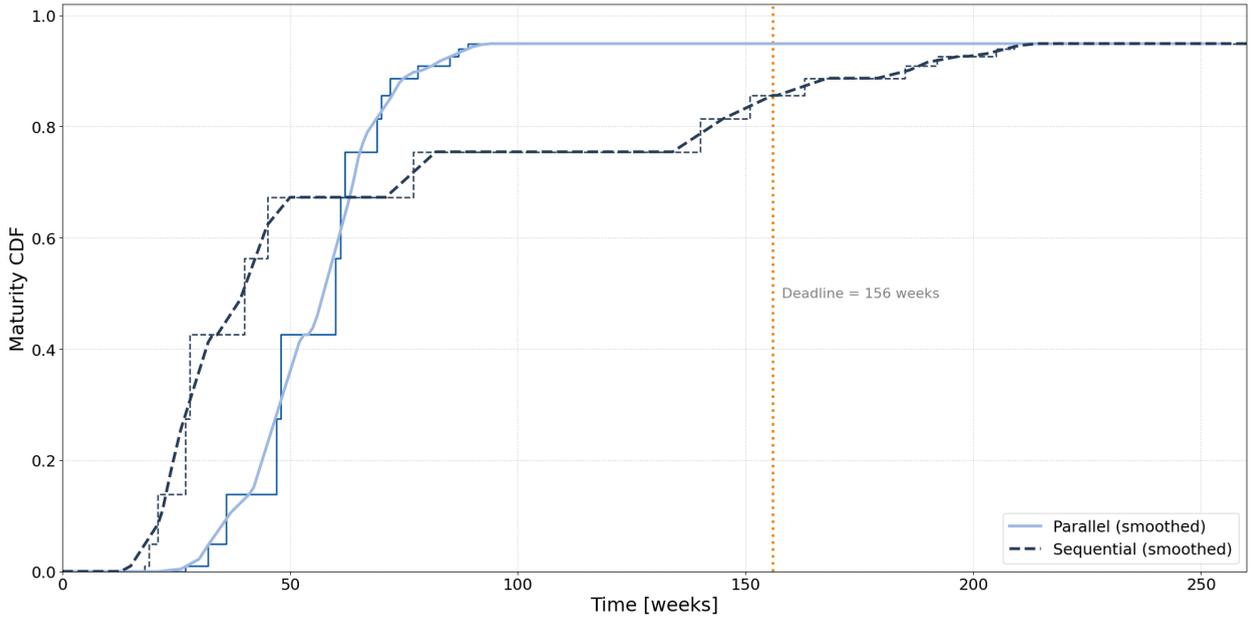

**Figure 3:**

Conversely, the sequential configuration introduces systemic constraints that fundamentally alter the emergent dynamics of the project. Although local performance indicators such as flow efficiency improve due to reduced rework, the overall project behaviour becomes brittle and temporally extended. The delayed crossing of the maturity CDF beyond the critical threshold demonstrates how precedence constraints amplify propagation delays and diminish the system's capacity to absorb variation. This rigidity limits adaptive responses and increases schedule sensitivity, revealing the inherent trade-off between control and agility within highly structured execution logics.

Furthermore, the probabilistic maturity CDF provides a valuable methodological perspective for interpreting the systemic consequences of execution structure under uncertainty. While the model does not explicitly include empirical TRL milestones, constructing the CDF from normalised task-completion trajectories enables the estimation of a probabilistic maturity envelope over time. Although this abstraction deviates from real-world R&D trajectories, it provides a meaningful approximation of how execution pathways impact the likelihood of achieving functional readiness within a defined planning horizon. The value of this approach lies not in predictive precision but in its ability to uncover how systemic characteristics—such as dependency structures, feedback loops, and rework propagation—shape the aggregate trajectory of innovation progress. In domains where timing is critical, such as compliance with regulatory milestones or stakeholder alignment, understanding how structural dynamics affect readiness probability can inform strategic scheduling, early-phase decision-making, and risk assessment.

It remains essential, however, to interpret these findings with appropriate epistemic caution. As outlined previously, the simulation framework is explicitly positioned as a proof-of-concept rather than a predictive or diagnostic instrument for operational use in industrial contexts. The model incorporates simplifying assumptions, most notably the exclusion of regulatory constraints and external shocks, as well as the use of synthetic parameterisation, which collectively limit its external validity. The simulated project structure represents a generic abstraction that does not capture the specificities of electrification technologies or sectoral processes. Consequently, the results concerning task dependencies should not be interpreted as forecasts of actual R&D trajectories but as illustrative insights into how architectural choices influence innovation dynamics under high uncertainty.

Despite these limitations, the internal coherence observed between model structure, execution dynamics, and emergent maturity behaviour provides confidence in the conceptual robustness of the approach. The hybrid AB–SD framework reproduces expected system patterns across different task configurations, generating plausible rework cycles, maturity lags, and agent-level bottlenecks without empirical calibration. This alignment between structural design and emergent behaviour confirms that the model operates consistently within its intended scope and achieves its broader methodological purpose: to enable simulation-based exploration of the systemic interdependencies shaping complex R&D project dynamics.

The finding that sequential project execution reduces rework task weeks by 88% and slightly decreases both client and quality management review times confirms that the hybrid AB–SD model behaves as expected. This outcome aligns with established System Dynamics theory on rework feedback loops, which posits that high task concurrency, as observed in the parallel scenario, amplifies the propagation of defects. When multiple tasks overlap, errors tend to be detected downstream, triggering cascading rework cycles (Lyneis and Ford, 2007). In contrast, sequential execution constrains error propagation within each task cycle, thereby reducing the accumulation of hidden rework and contributing to more stable process control.

### 3.1.2. Experiment 2 – Effect of team size

The second experiment examined how team size influences project progress and task completion within the hybrid AB–SD simulation framework. This analysis was motivated by the hypothesis that human resource availability represents a critical leverage point in R&D project execution, particularly when combined with different task-dependency structures. The experiment was designed to isolate the effect of workforce capacity under controlled conditions and to evaluate how varying levels of team size modulate system behaviour over time.

Two distinct task-execution logics were considered: (i) a parallel task-execution configuration and (ii) a sequential configuration, following the same structural setup described in Experiment 1. These configurations represent two archetypal operational modes in R&D workflows, capturing the contrast between concurrent and linearly dependent activities. The simulations were carried out for team sizes of one, two, three, four, five (base case), seven, and ten members. All other model parameters, as presented in Tables 5 and 6, were held constant across all experiments to ensure comparability. The underlying task structure, dependency matrix, and productivity-scaling rules were also kept unchanged.

The performance outcomes are summarised in Table 9, which reports total project duration, percentage of on-time task completion, and the rework ratio for each team size across both configurations. Figure 4 depicts cumulative task-completion trajectories for a representative subset of team sizes, while the full set of results is shown in Figure 8 in Appendix D. A comparative visual summary of project duration is also presented in Figure 5.

In the parallel execution configuration, a clear and consistent relationship emerges between team size and overall project performance. As team size increases, project duration decreases markedly and on-time completion rates improve. For instance, a project executed by a single team member required 377 weeks to reach full completion, with only 23.1% of tasks delivered within the 156-week deadline. Once the team reached four members or more, total project duration converged to approximately 88–89 weeks, achieving a 100% on-time completion rate. As shown in Figure 4(a), increasing the number of team members significantly accelerates task completion, particularly during the early phases of execution. This trend is reflected in the cumulative maturity curves presented in Appendix D, where team sizes of three, five, and ten display maturity values approaching 1.0 within the simulation horizon, whereas the single-member team exhibits a slower progression. Notably, none of the configurations achieved an exact maturity value of 1.0, reflecting the inherent stochastic variability in task execution and feedback cycles.

In the sequential execution configuration, where tasks must be completed in strict order, project performance varied substantially across different team sizes. Teams of one, two, four, and five members completed all 15 tasks within the 400-week simulation window. Among these, the five-member team achieved the shortest total project duration at 264 weeks, followed by the two-member team at 310 weeks, the one-member team at 318 weeks, and the four-member team at 350 weeks. Teams of three, seven, and ten members, by contrast, did not complete all tasks within the timeframe, reaching only 7, 11, and 9 completed tasks, respectively. As shown in Figure 8(b), the cumulative completion curves for these teams plateaued well before reaching task 15, indicating that execution halted before full completion. The bar charts in Figure 5(b) summarise these findings visually: solid bars represent fully completed projects, whereas diagonally hatched bars denote partial completion. None of the team sizes met the 156-week deadline, and only the five-member team completed the entire sequence in under 300 weeks.

| Team | Parallel Completion [Weeks] | Parallel On-time [%] | Parallel Rework Ratio [%] | Sequential Completion [Weeks] | Sequential On-time [%] | Sequential Rework Ratio [%] |
| --- | --- | --- | --- | --- | --- | --- |
| 1 | 377.0 | 23.08 | 2.3 | 318.00 | 60.00 | 1.79 |
| 2 | 333.0 | 66.7 | 0.08 | 310.00 | 73.33 | 0.15 |
| 3 | 220.0 | 93.3 | 1.13 | N/A | N/A | 0.33 |
| 4 | 88.0 | 100.0 | 0.24 | 350.00 | 53.33 | 0.29 |
| 5 | 88.0 | 100.0 | 1.24 | 264.00 | 66.67 | 0.19 |
| 7 | 80.0 | 100.0 | 0.17 | N/A | N/A | 0.15 |
| 10 | 89.0 | 100.0 | 0.12 | N/A | N/A | 0.12 |

**Table 9**
Team size influence on sequential task execution.

Further insights are provided in the maturity progression curves in Appendix 9, which show the cumulative development of maturity over time for selected team sizes. These curves confirm the previously observed completion patterns: teams of three and ten members exhibit trajectories that plateau well below full maturity, while smaller teams (one and two members) and the five-member team show continued, albeit gradual, progression. None of the configurations achieved a maturity probability of exactly 1.0 within the 400-week simulation horizon. Collectively, these results highlight the pronounced effect of team size on project outcomes under strict dependency constraints and underscore the importance of balancing workforce allocation with structural flexibility in R&D project design.

The results from Experiment 2 provide a detailed account of how team size influences project execution under both parallel and sequential task configurations within an R&D environment. In the parallel execution scenario (Figures 4 and D(a)), increasing team size consistently reduced total project duration and improved task completion rates within the 156-week deadline. A marked decrease was observed from 377 weeks for a single team member to approximately 88–89 weeks when team size reached four or more members. This behaviour reveals an apparent threshold effect: a minimum team size of four to five is required to fully exploit parallelism, beyond which additional resources provide diminishing returns. This threshold aligns with the resource-saturation phenomenon described in the System Dynamics literature, where increasing resources initially enhances performance but eventually becomes constrained by coordination overheads and diminishing marginal gains (Sterman, 2002). Although this result confirms the model's sensitivity to workforce capacity, the parallel configuration should be interpreted as an idealised boundary condition rather than a realistic depiction of R&D operations.

It remains improbable that a five-member team could complete a complex offshore R&D project, spanning early-phase development to commercialisation, within 156 weeks. This discrepancy suggests that the current parameterisation may overestimate agent productivity, highlighting a key avenue for future model calibration using empirical data.

In contrast, the sequential execution scenario exhibited more complex and less intuitive dynamics. Teams of three, seven, and ten members were unable to complete all fifteen tasks within the 400-week simulation window, reflecting inefficiencies arising from rigid task dependencies and allocation rules. These inefficiencies stem from the integer-based task distribution mechanism within the hybrid AB–SD model, as outlined in Section 2. Each team member sequentially claims tasks from an open queue and retains them through all stages—including quality management, client review, and potential rework—until completion. While this formulation realistically incorporates rework feedback loops, it also introduces scheduling bottlenecks when all agents are simultaneously occupied, preventing unassigned tasks from being initiated and causing premature stagnation in task progression.

These bottlenecks were particularly evident for teams of three, seven, and ten members, where uneven workload distribution resulted from the integer division of tasks among agents. For example, ten-member teams were assigned only one or two tasks per agent, leading to excessive waiting periods once the strict sequential dependencies were activated. Similar inefficiencies occurred in teams of three and seven, characterised by uneven workloads and extended idle times that ultimately prevented the initiation of final tasks. In all three cases, the alignment of task completion, review cycles, and agent availability led to situations where every agent remained occupied precisely when new tasks became available, thereby preventing their execution. This systematic under-completion persisted even when the simulation horizon was extended to 600 weeks.

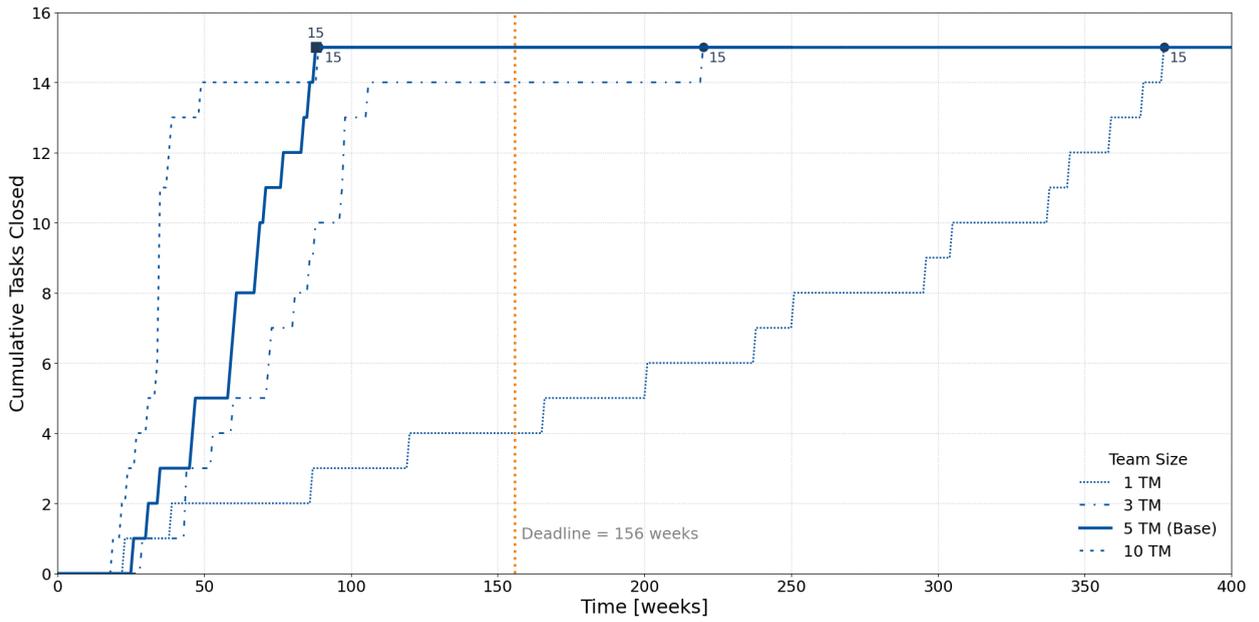

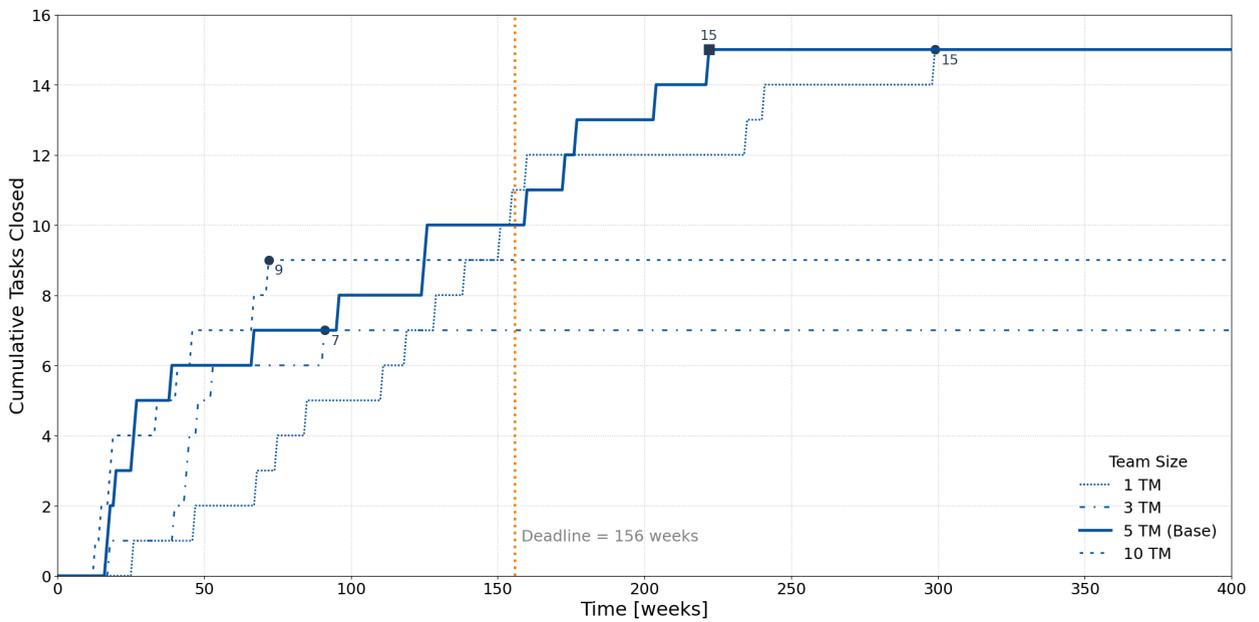

**Figure 4:** Cumulative number of closed tasks over time for selected team sizes under two execution configurations. Panel **(a)** shows the parallel configuration with independent tasks executed concurrently while panel **(b)** displays the sequential configuration, where tasks follow a fixed dependency order. The figure included a subset of team sizes to enhance visual clarity. The vertical orange line indicated the 156-week project deadline, and circle markers denote the point of full task completion.

These findings align closely with insights from the System Dynamics and Agent-Based Modelling literature, particularly the work of Shafieezadeh et al. (2020), who demonstrated that larger teams do not necessarily yield proportional performance improvements in tightly coupled project environments. Increased coordination complexity and longer waiting periods frequently offset potential productivity gains under strong sequential dependencies. The incomplete task execution observed in the simulations is therefore a predictable consequence of the structural constraints embedded in the hybrid model rather than a stochastic anomaly.

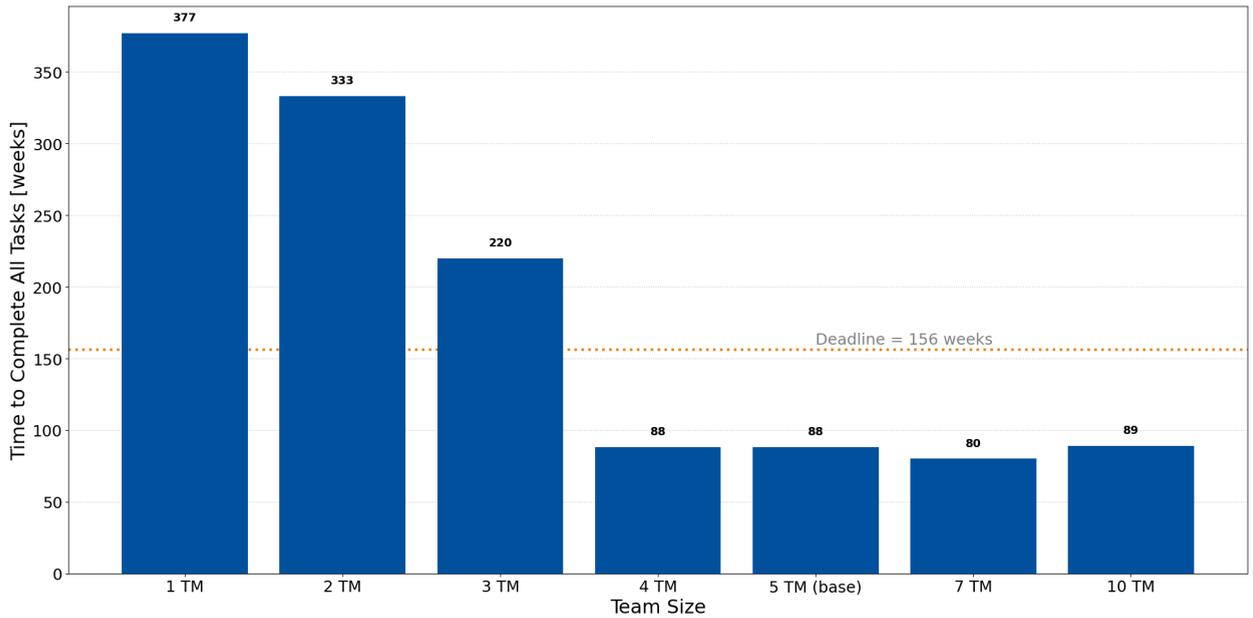

(a) Parallel

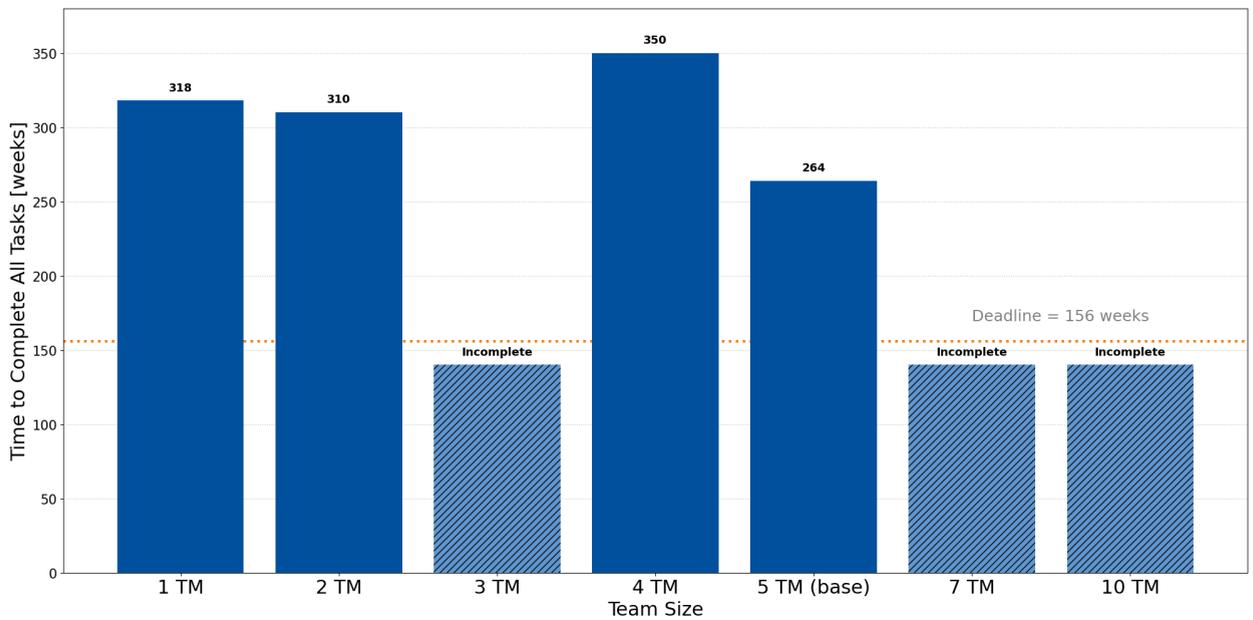

(b) Sequential

**Figure 5:** Total project duration for each time size under parallel **(a)** and sequential **(b)** task execution configurations. Bars represent the time required to complete all 15 tasks. Incomplete runs in the sequential case are shown with diagonally hatched bars, indicating that the project did not reach completion within the simulation window. The orange line marks the project deadline of 156 weeks.

Interestingly, this phenomenon did not occur in team configurations with one, two, four, or five members. These configurations naturally allowed at least one agent to become available at the precise moment when new tasks entered the queue, enabling continuous task progression and full project completion. This pattern suggests that the observed performance limitations primarily stem from discrete scheduling sensitivities rather than overall resource scarcity. The outcome underscores the crucial role of strategic task management in complex R&D systems, where even minor scheduling interactions can have significant systemic effects.

Although the hybrid AB–SD model effectively captures the intricate relationship between resource allocation, coordination, and task interdependency, it cannot yet be regarded as a definitive decision-making instrument. Nevertheless, the findings provide meaningful insights for R&D project management. They underline the importance of optimal team sizing, flexibility in task scheduling, and strategic coordination of dependencies. Implementing these practices can help mitigate the risks of delays and inefficiencies identified in the sequential-task configuration. While further model refinement and empirical validation are required to extend its predictive applicability, the current results already offer practical guidance for structuring team composition and managing interdependencies in complex R&D environments.

## 4. Conclusions

The main goal of this study was to develop and demonstrate a hybrid modelling framework capable of representing the complex dynamics of R&D project execution and its relationship to technological maturity. The model integrates the continuous feedback representation of System Dynamics with the decentralised perspective of Agent-Based Modelling, creating a unified structure for analysing the interaction between task-level behaviour, coordination mechanisms, and emergent project outcomes. This approach offers a methodological bridge between project execution dynamics and maturity assessment frameworks, providing a foundation for understanding how structural and behavioural factors jointly influence innovation progress.

The results confirm that the proposed framework reproduces the characteristic patterns of R&D system behaviour, including variability in throughput, rework propagation, and review-cycle delays. These outcomes reflect known empirical trends and theoretical principles from System Dynamics, where the degree of task coupling governs the balance between flexibility, stability, and time-to-readiness. The experiments demonstrate that parallel configurations facilitate faster progress and higher throughput, albeit at the expense of increased rework. In contrast, sequential structures reduce rework but extend the overall project duration and limit adaptability.

A further contribution of this work is the introduction of a probabilistic representation of technological maturity, based on cumulative distribution functions derived from simulated task completion. This feature enables a conceptual link between simulated execution and innovation metrics such as TRLs. Although the framework does not rely on empirical TRL milestones, it produces maturity trajectories that capture the probabilistic evolution of readiness under uncertainty, allowing early identification of potential risks and delays.

Within its proof-of-concept scope, the hybrid AB–SD model demonstrates internal coherence and conceptual robustness. The consistency observed between structure, behaviour, and emergent maturity trajectories indicates that the framework functions reliably within its defined boundaries. It serves as an exploratory and decision-support tool that facilitates the examination of alternative project architectures, investment priorities, and schedule strategies in R&D environments. The overall contribution lies in establishing a transparent, extensible and quantitative basis for simulation-based analysis of technological innovation, supporting the transition toward more data-informed and adaptive management of digital and industrial transformation processes.

## A. Software for Modeling

The implementation of the hybrid AB-SD simulation model presented in this paper was carried out using Python as the primary programming language. Python was chosen due to its widespread adoption in scientific computing, robust ecosystem of numerical libraries, and compatibility with various simulation frameworks. Specifically, the Business Prototyping Toolkit for Python (BPTK-Py) library was utilized, providing specialized tools and functionalities that supported the efficient and accurate simulation of the proposed model. The modular design facilitated by Python and Business Prototyping Toolkit for Python (BPTK-Py) significantly contributed to the iterative development and refinement of the simulation model.

### A.1. Conceptual Modeling Tool

The final implementation of the hybrid AB-SD model was carried out using the Business Prototyping Toolkit for Python (BPTK_Py), an open-source simulation framework developed by Transentis Labs (Transentis Consulting, 2024). BPTK_Py supports the integration of multiple simulation paradigms, such as SD, ABM, and DES, within a unified simulation loop. This made it particularly well-suited for modeling the co-evolution of micro-level agent behaviors and macro-level system feedback characteristic of R&D projects.

The framework offers a high-level interface for defining agent states, event-driven transitions, and differential equations, while also supporting real-time simulation, data logging, and post-processing by Python's scientific computing libraries (e.g., NumPy, Pandas, and Matplotlib) (Transentis GmbH, n.d.). These features allowed for seamless data analysis and visualization within the same development environment.

Key advantages of BPTK_Py include:

- A clear abstraction layer for defining agents, properties, and event-driven logic;
- Built-in collection of time series data across simulation runs;
- Built-in integration with Python's scientific computing stack (e.g., NumPy, Matplotlib, Pandas);
- Modular architecture that facilitates extensibility and experimentation with simulation logic.

Despite its flexibility, the adoption of BPTK_Py also introduced certain challenges. The framework imposes its own internal conventions and abstractions for managing time steps, state updates, and coupling between ABM and SD layers. This required a steep learning curve and non-trivial adaptation of modeling logic to align with the framework's execution engine. Furthermore, the available documentation and examples, while sufficient for introductory use, offered limited guidance on advanced hybrid integration, which occasionally slowed the implementation process. Nevertheless, BPTK_Py provides the essential capabilities required to realize the objective of this thesis. It enables the design of a hybrid simulation model that could integrate agent-level task execution dynamics with system-level workload evolution, thus supporting the exploratory analysis of complex R&D project behavior under uncertainty.

## B. Agent-Based Model Design

The agent-based component of the hybrid AB-SD model represents the micro-level structure of an R&D project. The model comprises three primary agent types, *Task Agent*, *Team-Member Agent*, and a *Controlling Agent*. These agents interact dynamically to simulate task execution, personnel assignment, learning effects, and project-wide coordination. The agent structure and interaction logic are inspired by modeling patterns demonstrated in the BPTK_Py documentation by Transentis Labs Transentis Consulting (2024). While adapted and extended for the context of offshore R&D projects, this provided a valuable foundation for structuring the task–developer dynamics within the ABM framework.

### B.1. Task Agent

In the agent-based layer of the hybrid model, *Task Agents* are used to represent individual R&D tasks. Unlike SD approaches, where such tasks might be defined via index-based stock arrays (e.g., in tools like Vensim), the ABM approach allows explicit modeling of task heterogeneity and dynamic interactions without requiring expensive SD simulation software or complex vectorization. To represent the diverse and interdependent nature of R&D activities, the simulation deploys multiple instances of the task agent, where each instance corresponds to a unique task characterized by its effort, dependencies, and execution dynamics.

| Attribute | Description | Unit |
| --- | --- | --- |
| *State* | | |
| `open` | Task is ready to be picked up but work has not yet started | [-] |
| `waiting.dependencies` | Task cannot start until all predecessor tasks have been completed | [-] |
| `in.progress` | A developer is actively working on the task | [-] |
| `qm.review` | Task is undergoing internal quality management review | [-] |
| `client.review` | Task is being reviewed by a client or project lead for external validation | [-] |
| `rework` | Task failed client or quality review and requires partial re-execution | [-] |
| `closed` | Task has passed all reviews and is fully completed | [-] |
| *Properties* | | |
| `effort` | Initial total workload required to complete task | [Work units] |
| `remaining.effort` | Remaining work left to complete the task | [Work units] |
| `dependencies` | List of task IDs, list of predecessor task IDs | [-] |
| `rework probability` | Probability that the task will require rework | [%] |
| `qm.delay` | Duration of the internal quality assessment process | [Weeks] |
| `client.delay` | Duration of the external client review process | [Weeks] |
| `assigned` | Boolean indicating whether the task is currently assigned to a team-member (true/false) | [-] |

**Table 10**
Underlying state and properties for the Task Agent in the ABM model structure.

Each task follows a defined lifecycle through a series of discrete states. A task agent is initially created in the `open` state but remains inactive until all its predecessors are resolved. If dependencies are present and not yet fulfilled, the task enters a temporary queue state, `waiting.dependencies`. Once all prerequisites are satisfied, the task becomes eligible for assignment and, upon being picked up by a team member, transitions into `in.progress`.

Work on the task proceeds according to the effort input the assigned team member provides. Upon completing the required workload, the task proceeds to an internal quality assurance review, `qm.review`, followed by an external client validation phase, `client.review`. Based on a probabilistic outcome, the task may either be approved and closed or returned for partial re-execution in the `rework` state. In such cases, a portion of the original workload is reassigned, simulating the real-world cycles that often occur during early-stage technology development/ R&D projects.

A complete overview of the task agent's states and properties in the ABM framework is provided in Table 10, which outlines the variables that govern each task agent's behavior and lifecycle. The underlying assumptions and parameterization strategies for these variables are further described in Section 2.6.

### B.2. Team Member Agent

The *Team Member* agent represents an individual developer or engineer within the R&D project. Each agent operates autonomously, selecting tasks, performing technical and research work, and learning from experience/improving its productivity through cumulative learning, a process that reflects the acquisition of domain knowledge and task-specific competencies over time.

At any given point, a team member is in one of two states: `available` or `busy`. When available, the agent scans all task instances of the task agent for eligible assignments, specifically those in `open` or `rework` states that are not currently assigned and whose dependencies are fully resolved. Upon assignment, the team member transitions to the busy state and initiates task execution by triggering a `task.started` event in the corresponding task agent.

| Attribute | Description | Unit |
|---|---|---|
| *State* | | |
| `available` | Team member is idle and ready to take on a new task | [-] |
| `busy` | Team member is currently working on an assigned task | [-] |
| *Properties* | | |
| `task` | Reference to the currently assigned task, or None if unassigned | [Work units] |
| `completed.task` | Total number of tasks completed by the agent | [Count] |
| `learning.factor` | Sensitivity of productivity growth to task completion | [-] |
| `max.productivity` | Maximum personal productivity the agent can achieve | [-] |
| `personal.productivity` | Current productivity level based on cumulative learning | [-] |

**Table 11**
Underlying state and properties for the *Team Member* agent in the ABM model structure.

Work is performed incrementally during each simulation timestep, and the amount of effort contributed is governed by the agent's effective productivity, defined as the product of a global productivity factor and the agent's individual productivity level. At the same time, the global productivity factor is intended to reflect the system-wide effects, such as schedule pressure and workload intensity derived from the SD layer. Individual productivity evolves dynamically during the simulation. It is modelled using an S-curve learning function, which models the principle of experience-driven capability growth. Productivity increases as the agent completes more tasks, but follows a diminishing returns curve, capturing the realistic ceiling in performance improvement.

### B.3. Controlling Agent

The *Controlling* agent is a centralized coordination mechanism in the hybrid model, acting as the interface between the ABM and the SD layer. Its primary function is to monitor and expose key global parameters, specifically `productivity` and `schedule.pressure`. These parameters influence agent behavior across the model. Additionally, it accumulates task-level workload data for synchronization with the SD model and could track the TRL trajectory for the R&D project in a more extended implementation.

Upon initialization, the agent sets baseline values for `productivity` and `schedule.pressure`, which conceptually represent the system-wide dynamics. The variables are registered as agent properties using the simulation engine's property system, enabling them to be accessed globally by other agents (e.g., team members) during simulation execution.

In addition, the *Controlling* agent aggregates task-level effort data. Specifically, it computes the total remaining effort across all *Task* agents. These aggregates are passed to the SD model as inputs (`abm.open.effort`, `abm.in.progress.effort`, and `abm.rework.effort`). This setup allows the SD to adapt in real time to shifts in task distributions.

A further key functionality of the *Controlling* agent is the dynamic estimation of technological maturity through a cumulative distribution function (CDF). The mathematical formulation and interpretation of the CDF are described in detail in Section 2.4. This CDF-based estimation supports the evaluation of technological progress over time and offers an interpretable maturity index.

| Team Size | Completion Time [Weeks] | On-time Completion [%] | Flow Efficiency [%] |
|---|---|---|---|
| 1 | 377.0 | 26.7 | 10.7 |
| 2 | 333.0 | 66.7 | 60.6 |
| 3 | 220.0 | 93.3 | 33.9 |
| 4 | 88.0 | 100.0 | 36.4 |
| 5 | 88.0 | 100.0 | 48.1 |
| 7 | 80.0 | 100.0 | 59.5 |
| 10 | 89.0 | 100.0 | 79.1 |

**Table 12**

Team size influence on parallel task execution

| Team Size | Completion Time [Weeks] | On-time Completion [%] | Flow Efficiency [%] |
|---|---|---|---|
| 1 | 299.0 | 73.3 | 10.7 |
| 2 | 329.0 | 60.0 | 60.6 |
| 3 | 91.0 | 100.0 | 33.9 |
| 4 | 341.0 | 53.3 | 36.4 |
| 5 | 222.0 | 66.7 | 48.1 |
| 7 | 123.0 | 100.0 | 59.5 |
| 10 | 72.0 | 100.0 | 79.1 |

**Table 13**

Team size influence on sequential task execution

| Metric (SD) | Parallel | Sequential | Unit |
|---|---|---|---|
| $WTD_{avg}$ | 465.22 | 39.40 | Work Units |
| $WQA_{avg}$ | 7.29 | 15.95 | Work Units |
| $WCR_{avg}$ | 211.16 | 810.72 | Work Units |
| $WA$ | 138.06 | 163.30 | Work Units |
| $WIR_{tot}$ | 352.00 | 72.00 | Work Units/Week |
| $WR_{tot}$ | 268.43 | 249.80 | Work Units/Week |
| Rework Ratio | 121.93 | 44.15 | % |

**Table 14**

.

## C. Experiment 1

## D. Experiment 2 - Effect of Team Size on Project Execution

The list below outlines how the 15 project tasks were distributed among agents for each team size. The allocation followed an even distribution approach, with any remainder tasks assigned to a subset of agents.

- For a team of 1 member, all 15 tasks are assigned to the single agent.
- For a team of 2 members, each agent is assigned 7 tasks, with one agent receiving an additional task (8 tasks).
- For a team of 3 members, each agent is assigned exactly 5 tasks.

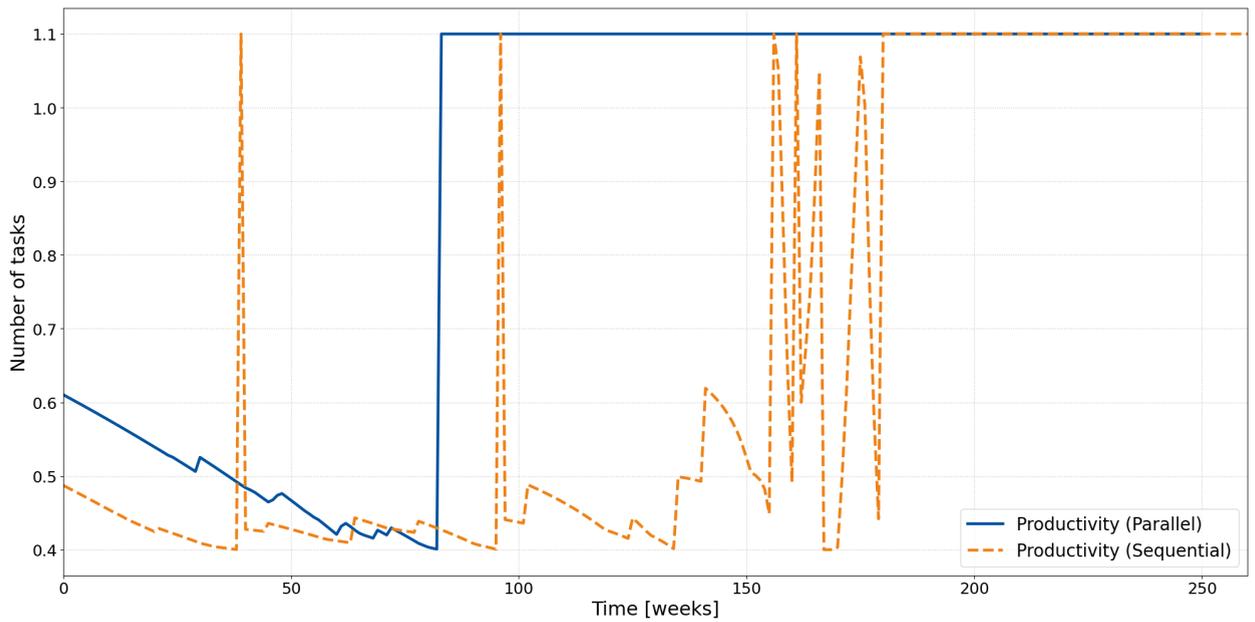

(a) Productivity for parallel (blue) and sequential (orange) configurations.

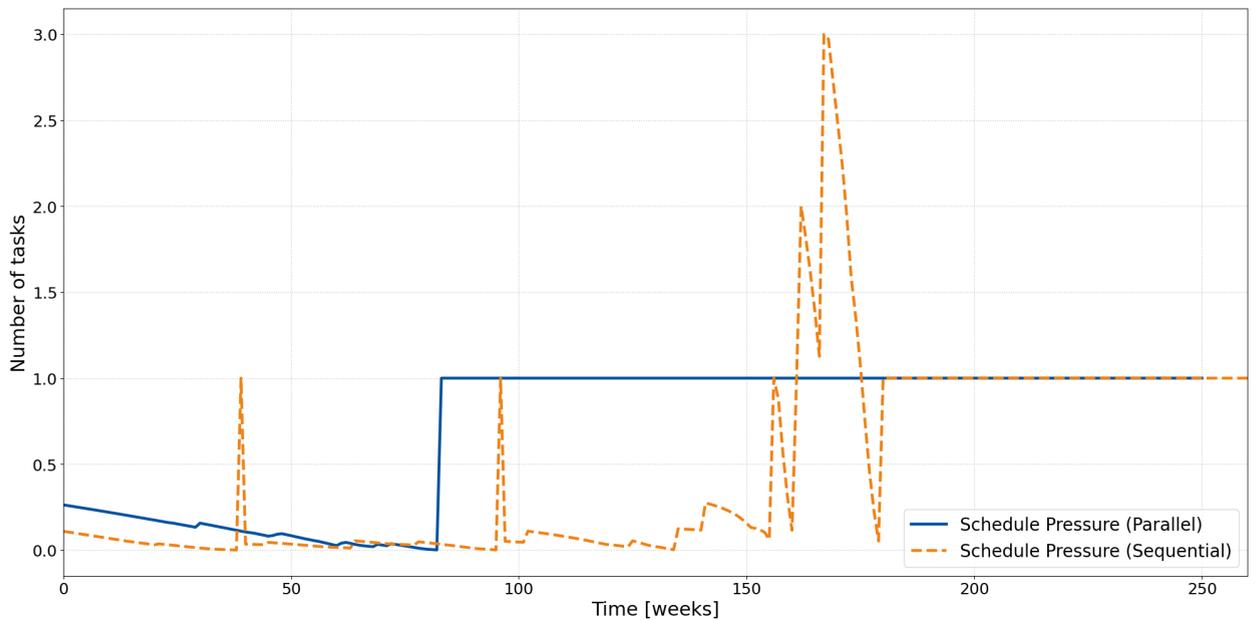

(b) Schedule pressure for parallel (blue) and sequential (orange) configurations.

**Figure 6:** Performance metrics for Experiment 1, two configurations of dependency: parallel and sequential. **(a)**, displays productivity and **(b)**, represent schedule pressure.

- For a team of 4 members, three agents are assigned 4 tasks each, while one agent receives 3 tasks.
- For a team of 5 members, each agent is assigned exactly 3 tasks.
- For a team of 7 members, six agents are assigned 2 tasks each, and one agent receives 3 tasks.
- For a team of 10 members, five agents are assigned 2 tasks each, and five agents receive 1 task each.

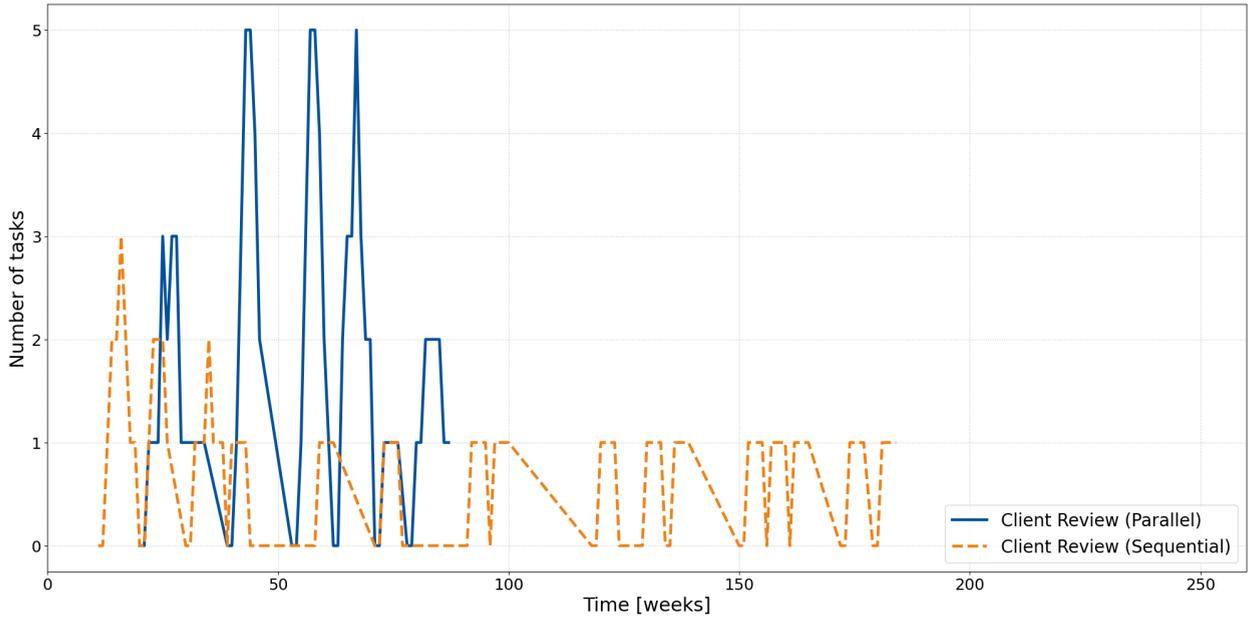

(a) Client review for parallel (blue) and sequential (orange) configurations.

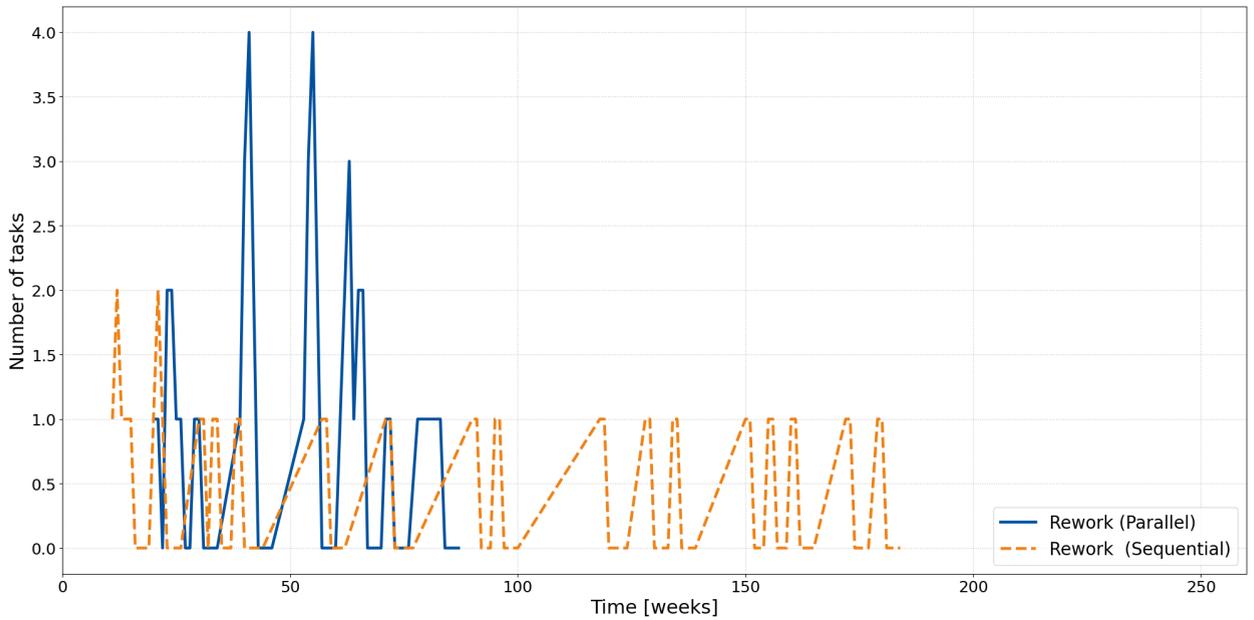

(b) Rework for parallel (blue) and sequential (orange) configurations.

**Figure 7:** Performance metrics for Experiment 1, two configurations of dependency: parallel and sequential. **(a)**, displays client review and **(b)**, represent rework.

## E. TRL

## Acknowledgement

This work was done in the framework of Project 101119358, 'PROSAFE', funded by the Marie Skłodowska-Curie Actions programme, HORIZONMSCA-2022-DN-01.

ABM-SD hybrid framework applied to R&D

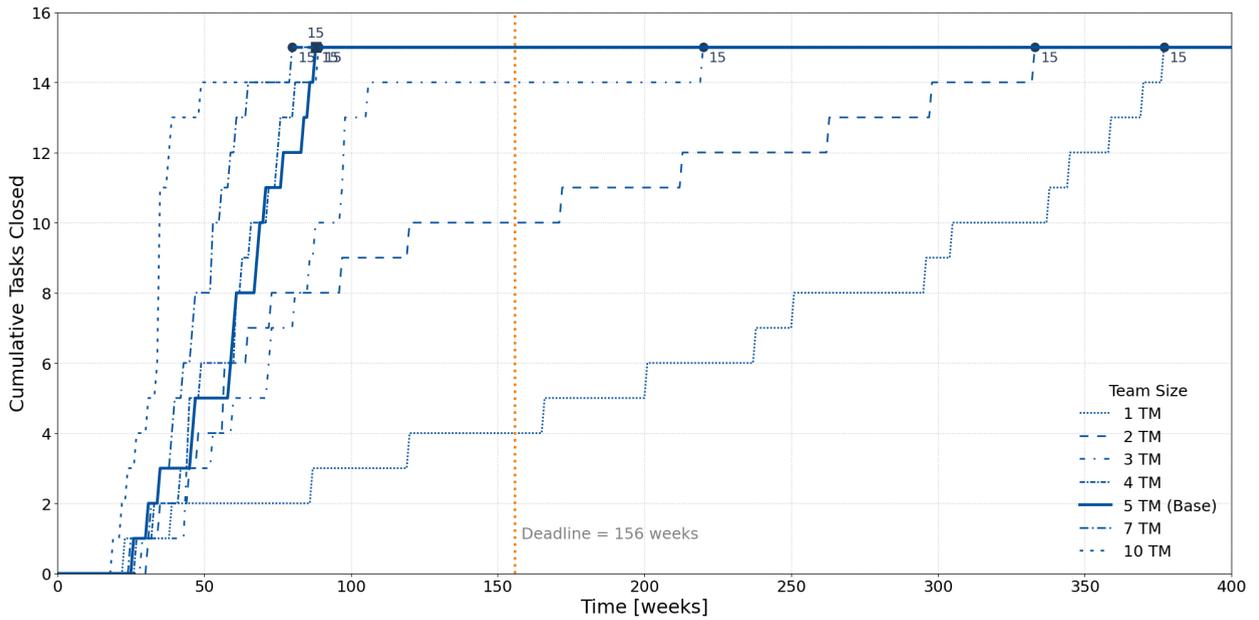

(a) Parallel

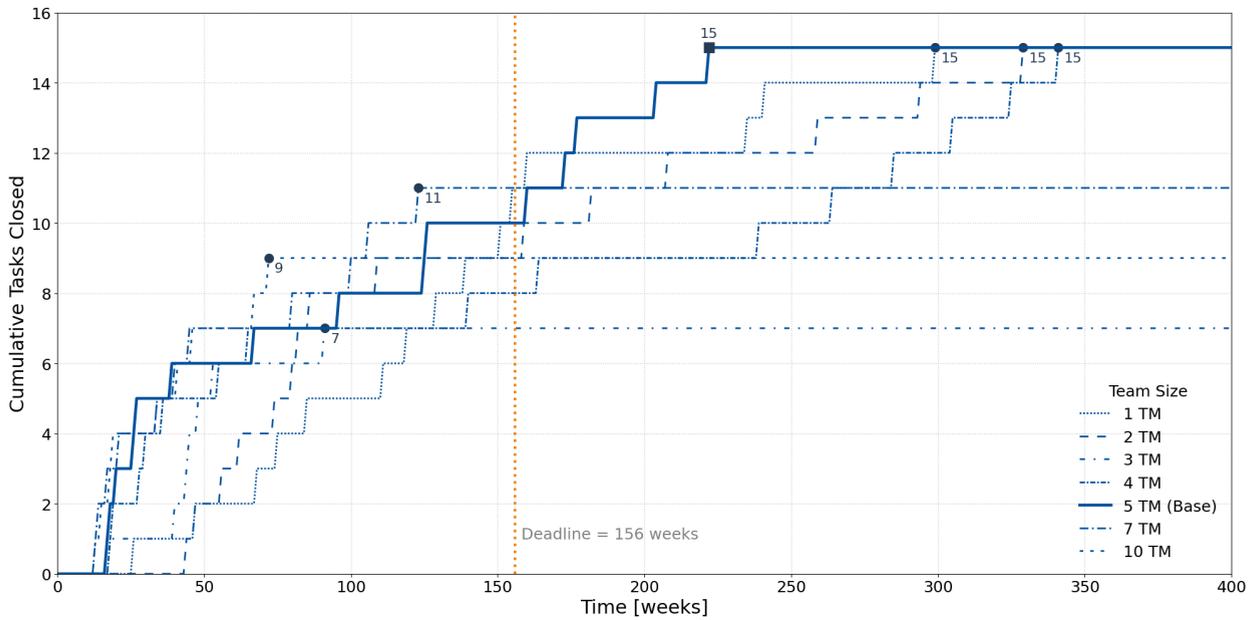

(b) Sequential

Figure 8: Cumulative number of closed tasks over time for different team sizes under two execution structures. **(a)** shows the parallel configuration, in which all tasks are independent and may be executed concurrently, whereas **(b)** represents the sequential configuration, where tasks are constrained by strict dependencies and executed in a fixed order. Each curve represents a simulation run for a specific team size: one, two, three, four, five (base case), seven, and ten team members. The vertical orange line indicates the project deadline, which is 156 weeks. Circle markers denote the completion time at which all 15 tasks are closed for each configuration. These plots illustrate how team capacity interacts with task structure to influence the timing and pacing of work completion.

ABM-SD hybrid framework applied to R&D

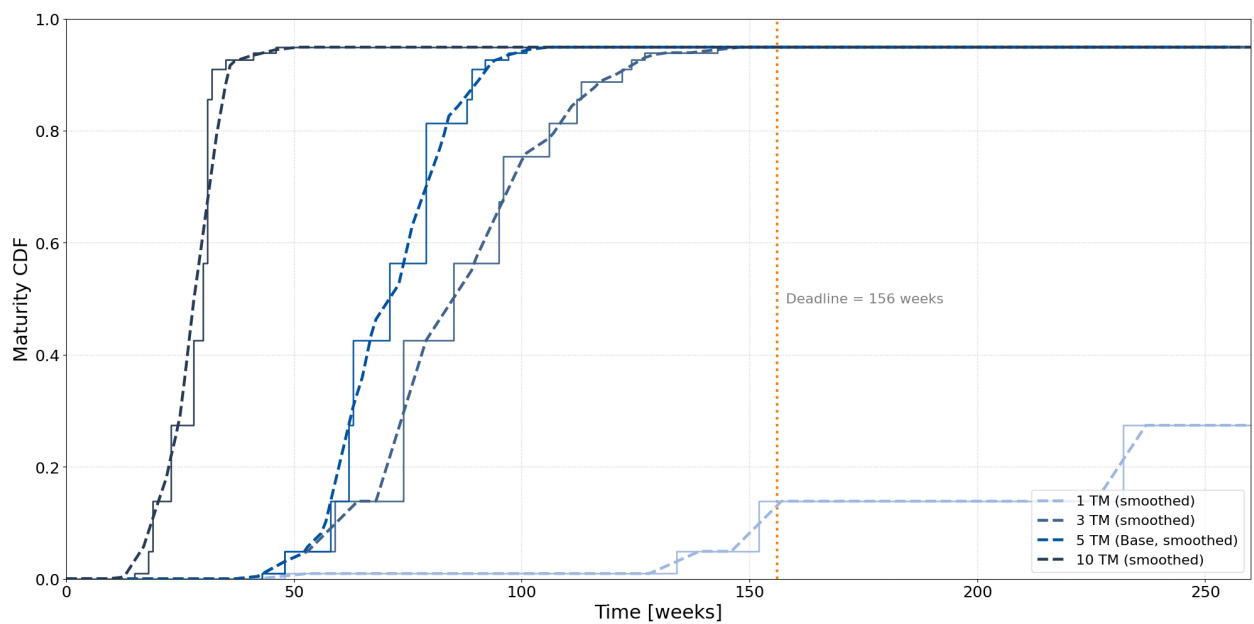

(a) Parallel

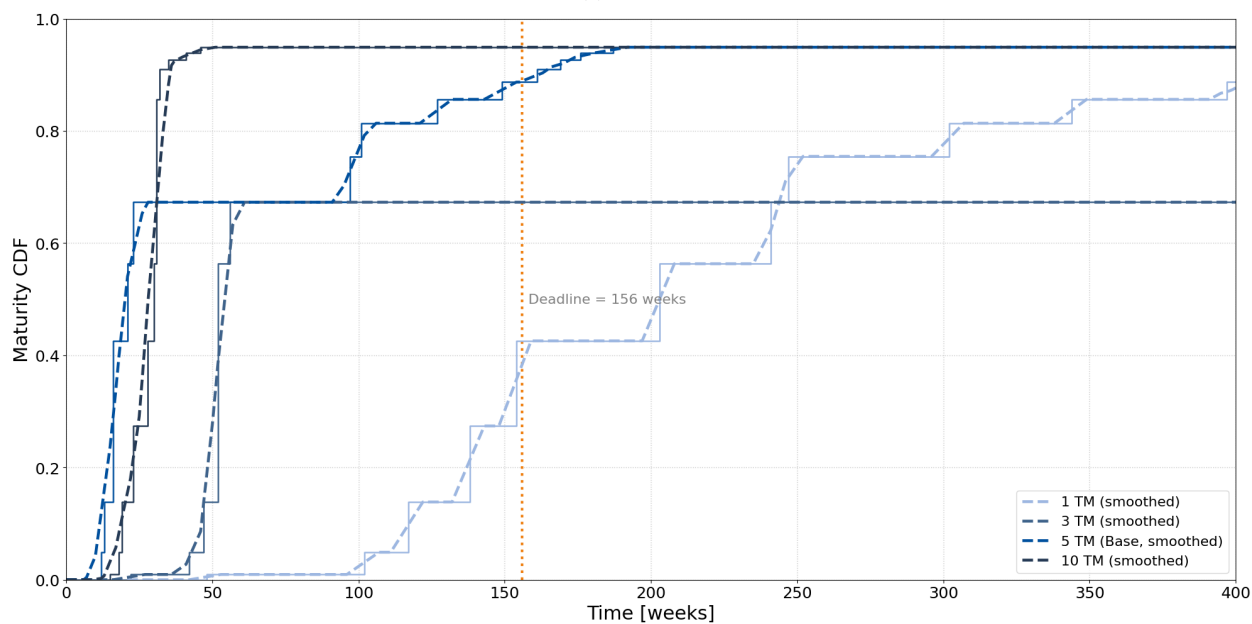

(b) Sequential

**Figure 9:** Cumulative maturity progression over time for selected team sizes in **(a)** parallel and **(b)** sequentital task configurations. Curves represent smoothed maturity trajectories (CDF) for one, three, five, and ten team members. In the parallel case **(a)**, maturity increases rapidly with larger team sizes, converging toward full completion before the deadline. In contrast, the sequential case **(c)** exhibits delayed and plateaued maturity trajectories, particularly for team sizes of three and ten.

| TRL | Definitions, NASA | Definitions, Buchner et. al Buchner et al. (2018, 2019) |
|---|---|---|
| 1 | Basic principles observed and reported | Potential applications identified; conceptual design initiated based on scientific principles. |
| 2 | Technology concept and/or application formulated | Concept elaborated; preliminary performance demonstrated via analytical and experimental models. R&D activities planned. |
| 3 | Analytical and experimental critical function and/or characteristic proof-of-concept | Functional laboratory tests of prototype; qualitative validation of expected reaction or process behavior. |
| 4 | Component and breadboard validation in laboratory environment | Laboratory validation of functionality under controlled conditions; preliminary process development. |
| 5 | Component and breadboard validation in relevant environment | Controlled laboratory validation of components; scale-up to pilot level begins. |
| 6 | System/subsystem model or prototype demonstration in a relevant environment | Low-rate pilot plant operated; detailed process model derived; feasibility of integration assessed. |
| 7 | System prototype demonstration in the planned operational environment | Pilot plant performance optimized; equipment specifications validated for full-scale transfer. |
| 8 | Actual system completed and qualified through test and demonstration in the operational environment | Products/processes integrated into organizational structure; full-scale plant constructed. |
| 9 | Actual system proven through successful system and/or mission operations | Industrial-scale audited operations; enforceable performance guarantees confirmed. |

**Table 15**
Comparison between the NASA TRL framework Mankins (2009) and the adapted TRL framework for chemical process industries, fragment from the study of Buchner et al. (2018, 2019).